# AI Security and Cyber Risk in IoT Systems

**Petar Radanliev[1,2*], David De Roure[3], Carsten Maple[4], Jason R.C. Nurse[5], Razvan Nicolescu[6], Uchenna Ani[7]**

[1]Department of Computer Sciences, University of Oxford;

[2] Department of Computer Science, School of Computing and Engineering, Huddersfield University, Huddersfield, United Kingdom

[3]Oxford e-Research Centre, Department of Engineering Sciences, University of Oxford,

[4]WMG Cyber Security Centre, University of Warwick,

[5]School of Computing, University of Kent, UK,

[6]University College London, UK,

[7]School of Computer Science and Mathematics, Keele University

*Corresponding author: Petar Radanliev[1*]: petar.radanliev@cs.ox.ac.uk*

**Funding:** This work has been supported by UK EPSRC under grant number EP/S035362/1.

**Acknowledgements:** Eternal gratitude to the Fulbright Visiting Scholar Programme.

**Conflicts of Interest:** The authors declare no conflict of interest.

**ABSTRACT:** Internet-of-Things (IoT) refers to low-memory connected devices used in various new technologies, including drones, autonomous machines, and robotics. The article aims to understand better cyber risks in low-memory devices and the challenges in IoT risk management. The article includes a critical reflection on current risk methods and their level of appropriateness for IoT. We present a dependency model tailored in context towards current challenges in data strategies and make recommendations for the cybersecurity community. The model can be used for cyber risk estimation and assessment and generic risk impact assessment. The model is developed for cyber risk insurance for new technologies (e.g., drones, robots). Still, practitioners can apply it to estimate and assess cyber risks in organisations and enterprises.

Furthermore, this paper critically discusses why risk assessment and management are crucial in this domain and what open questions on IoT risk assessment and risk management remain areas for further research. The paper then presents a more holistic understanding of cyber risks in the IoT. We explain how the industry can use new risk assessment, and management approaches to deal with the challenges posed by emerging IoT cyber risks. We explain how these approaches influence policy on cyber risk and data strategy. We also present a new approach for cyber risk assessment that incorporates IoT risks through dependency modelling. The paper describes why this approach is well suited to estimate IoT risks.

**INDEX TERMS** Artificial Intelligence; Internet-of-Things; cyber risk management; cyber risk assessment; cyber risk estimation; cyber risk insurance; risk impact assessment.

# 1 Introduction - Defining the Internet-of-Things (IoT)

The fast Internet of Things (IoT) adoption has transformed modern industries and daily life, creating interconnected environments that deliver unprecedented efficiency and convenience. However, this extensive integration of IoT devices has also introduced significant cybersecurity risks. The Internet of Things (IoT) has attracted the attention of cybersecurity professionals after cyber-attackers started using IoT devices as botnets (Palekar and Radhika 2022). IoT devices are often vulnerable to various cyber threats, including distributed denial-of-service (DDoS) attacks, botnet exploitation, and data breaches, all of which can compromise critical systems' integrity, confidentiality, and availability. Understanding and mitigating the risks associated with IoT deployments is crucial in this evolving landscape, especially given the interdependencies between IoT components and systems.

## *1.1 Motivation*

The primary motivation for this research arises from the pressing need to develop a comprehensive framework for assessing cyber risks in IoT environments. While several risk assessment models have been developed, only some fully capture the unique dependencies and interactions between IoT devices, networks, and services. These dependencies introduce cascading risks, where vulnerabilities in one component can propagate through the entire system, amplifying the impact of an attack.

Additionally, existing risk models often need more real-time adaptability and need to consider the heterogeneity of IoT systems, where devices from different manufacturers and platforms interact in dynamic and unpredictable ways. Given the increasing reliance on IoT in critical sectors such as healthcare, industrial automation, and smart cities, a more robust, adaptable, and scalable risk assessment model is urgently needed.

The research also addresses the gap in effectively utilising AI/ML techniques for real-time risk assessment in IoT environments while ensuring these models are explainable and transparent to decision-makers. This is particularly important for building trust in AI-driven cybersecurity solutions and ensuring their alignment with organisational goals.

### 1.2 Contributions

This paper makes the following key contributions:

1. **A Dependency-Based Cyber Risk Model for IoT Systems**: We propose a novel dependency-based risk assessment framework that captures the interdependencies between IoT components and their cascading effects on overall system security. The model systematically quantifies and mitigates risks based on the interaction between devices, networks, and services.
2. **Incorporation of AI/ML for Dynamic Risk Estimation**: The proposed model integrates AI/ML techniques to enable real-time risk assessment. The machine learning models are trained on diverse data sources, including network traffic, vulnerability databases, and incident logs, to predict and prioritise risks in dynamic IoT environments. Explainable AI (XAI) ensures that these predictions are transparent and interpretable to cybersecurity professionals.
3. **Generalisation of the Risk Framework Across IoT Domains**: We demonstrate the applicability of the proposed model across various IoT domains, including smart cities, healthcare, and industrial IoT. The framework is adaptable to different types of IoT systems, regardless of device heterogeneity or scale, making it a versatile tool for risk assessment in diverse settings.
4. **Integration of Risk Transference Strategies**: This research explores risk transference mechanisms, such as cyber insurance and third-party liability agreements, to mitigate IoT cyber risks' financial and operational impact. We discuss how these strategies can be effectively implemented within the proposed framework.
5. **Empirical Validation Using the BoT-IoT Dataset**: The proposed model is validated using the BoT-IoT dataset, a comprehensive and realistic representation of IoT network traffic and attack scenarios. We provide an in-depth analysis of the model's performance in detecting and mitigating risks, and we compare it with alternative risk assessment frameworks to highlight its effectiveness.

IoT-based cyber-attacks often take the form of distributed denial of service (DDoS) attacks, where the attacker uses the hacked IoT devices as clones to infect or stop operations in other parts of the network. Various cybersecurity solutions have been proposed, including *'deep learning based malicious behaviour detection in cloud computing'* (Bhingarkar et al. 2022), *'sensing and detection algorithms'* (Zhang 2021) and the *'intelligent warehouse monitoring based on distributed system and edge computing'* (Lin et al. 2021). IoT is defined as networked objects communicating data between networks and humans ("The PETRAS National Centre of Excellence – PETRAS" 2022). The development of IoT has provided opportunities for social and economic interaction in many areas, such as supply chain management, social media, medicine, and energy consumption (for example, smart health devices). IoT employs sensors and actuators and applies to various protocols, domains, and applications, e.g. cyber-physical systems, technologies related to smart grids, smart homes, intelligent transportation and smart cities. Some technologies used daily are currently not connected to the Internet, such as gas meters, house lights, healthcare devices, water distribution systems, cars, and other road transport vehicles. However, such devices are increasingly becoming digitally connected and can communicate through mobile (or wireless) networks, e.g., connected spaces, smart meters and autonomous cars. Ultimately, IoT may revolutionise the existing business ecosystem because connected objects can reduce costs, optimise processes, and enable new business models by automating data flow, centralised processing of data, and intelligent use of the data.

With the increased relevance of IoT for business, cyber security importance is growing (Pigman 2019) and there are increasing security and privacy challenges (Maras and Wandt 2019). New technologies also come with new risks (Constance 2017) that traditional risk assessment/management methods have not anticipated or predicted (Crawford and Sherman 2018). It has been argued that quantitative risk assessments do not necessarily offer a unique rationality that pinpoints a single right course of action but rather probabilities that require moral assessment for action (Adams 1995). This kind of assessment can vary across domains and populations. For example, in financial markets, the complexity aspect of risk is of major importance. In contrast, in consumer markets, people are increasingly trained and habituated to act in the present regarding future risks (Caplan 2000).

Different cyber risk valuation models have emerged recently, including a model based on *'computationally efficient solution.. operating under the probable impact of typed cyber-physical attacks'* (Kovtun, Izonin, and Gregus 2022), or applying deep learning to detect *'Trojan malware in bio-cyber attacks'* (Islam et al. 2022). However, in evaluating the impact of risk, conventionally, it is considered, essentially, that Risk = Likelihood × Consequences. However, we do not have probabilistic data on the likelihood or consequences, and without such data, the industry's understanding of IoT cyber risk is still in its infancy (Aggarwal and Reddie 2018). Empirical results have found that the aggregate frequency of data breaches is stable over time (Edwards, Hofmeyr, and Forrest 2016; Wheatley, Maillart, and Sornette 2016). Still, future attacks are expected to increase (Leverett and Kaplan 2017) with IoT systems and other digital infrastructure. Digital expansion also changes the cyber risk profile, making it difficult to quantify with historical measures. In addition, there are no standards, regulations, or policies on risk assessment that measure the impact, cost, and probabilities of cyber-attacks in specific IoT systems (Srinivas, Das, and Kumar 2019). For example, if we consider cyber risk in general, the estimates of impact range from 300bn and $1tn (Biener, Eling, and Wirfs 2014), $400bn to over $575bn (DiMase et al. 2015), or $400bn to over $2tn (Shackelford 2016a). Although these figures could be correct in the parameters of the assessments, the difference presents a rationale for some literature to argue that the real impact of cyber risk is unknown (Shackelford 2016b). This motivates our attempt to define a process for standardising a unified cyber risk assessment approach.

In an IoT context, the most challenging aspects are risk's dynamic and complex aspects, including assessment of safety and security, co-existing of different producers and vintages, common cause failures and cascading risks.  Although, like cyber risk in general, IoT risk can be decomposed into different risk verticals. For example, because of the low cost of IoT devices, it is generally assumed that IoT devices cannot be adequately secured and, therefore, logical placement of security capability is in the communications network (Anthi, Williams, and Burnap 2018). To emphasise these differences, this paper articulates some of the possible security risks in the communications network, particularly the risk from distributed ownership and control of IoT systems. To develop and test the new approach for cyber risk estimation and assessment, in this study we used the *'BoT-IoT'* dataset[1], designed by the Cyber Range Lab of UNSW Canberra Cyber.

### *1.3 Justification for the Use of the BoT-IoT Dataset*

The BoT-IoT dataset was chosen for this study due to its comprehensive and realistic representation of IoT network traffic, which includes a wide range of attack scenarios. Developed by the Cyber Range Lab of UNSW Canberra, this dataset is designed explicitly for IoT environments. It includes various simulated attacks such as distributed denial-of-service (DDoS), keylogging, data theft, and information gathering. The dataset's diversity in attack types and network traffic allows for a holistic analysis of IoT-related cyber risks, particularly in botnet-driven attacks, which are among the most prevalent in IoT systems.

Moreover, the BoT-IoT dataset offers the following advantages:

- **Realistic Traffic Simulation**: The dataset captures real-world IoT traffic patterns, making it highly suitable for testing intrusion detection and risk assessment methods in heterogeneous IoT environments.
- **Diverse Attack Vectors**: It includes multiple attacks, such as DDoS, brute force, and OS and service scanning, relevant to understanding a wide array of IoT cyber risks.
- **Detailed Labelling**: The dataset is labelled, allowing for supervised machine learning approaches in identifying and mitigating threats, which is crucial for assessing the effectiveness of AI-based risk assessment models.

## 2 Alternative Datasets

[1] https://ieee-dataport.org/documents/bot-iot-dataset

Several alternative datasets could have been considered for this study, though they have certain limitations compared to BoT-IoT. These include:

1. **Kitsune Dataset**: This dataset focuses on the network traffic of IoT devices and has been widely used in anomaly detection. However, it is more limited in terms of attack variety and lacks certain botnet-specific data that is crucial for understanding large-scale distributed IoT attacks.
2. **TON_IoT Dataset**: Another comprehensive dataset developed by UNSW Canberra, the TON_IoT dataset contains IoT telemetry data, network traffic, and operating system logs. While useful, it is geared more toward industrial IoT (IIoT) environments and does not focus as heavily on botnet behaviour, which is the primary threat discussed in this paper.
3. **IoT-23 Dataset**: This dataset provides labelled IoT traffic data with malware analysis, but it is more focused on malware rather than the broad spectrum of cyber risks in IoT environments, making it less suitable for this study's goals.

While other datasets exist, the BoT-IoT dataset was chosen for its relevance to the focus of this study (evaluating the risks of IoT-based botnet attacks) and for its detailed attack scenarios that allow for robust risk estimation and assessment.

### *2.1 Organisation of the Paper*

The rest of this paper is organised as follows:

- **Section 2: Background and Related Work**: This section reviews the current state of IoT cybersecurity, including existing risk assessment models and their limitations. We also discuss the use of AI/ML in cybersecurity and highlight the gap that this research addresses.
- **Section 3: Proposed Dependency-Based Risk Assessment Model**: In this section, we detail the proposed model, explaining the methodology behind dependency analysis, the incorporation of AI/ML, and the use of explainable AI techniques. We also provide a formal definition of the risk estimation process.

- **Section 4: Data Sources and AI/ML Implementation**: This section describes the data sources used for training the machine learning models, including network traffic, device telemetry, vulnerability databases, and external threat intelligence. We explain the model's architecture and the machine-learning techniques employed.
- **Section 5: Empirical Evaluation and Results**: Here, we present the results of the empirical validation using the BoT-IoT dataset. We compare the performance of the proposed model against existing frameworks and discuss its effectiveness in detecting and mitigating IoT-related cyber risks.
- **Section 6: Discussion and Generalisation**: This section discusses the generalisability of the proposed model across various IoT domains. We provide case studies in healthcare, smart cities, and industrial IoT to demonstrate its broad applicability.
- **Section 7: Conclusion and Future Work**: We conclude the paper by summarising the key findings and outlining potential areas for future research, particularly in refining the AI/ML techniques and further validating the model in live IoT environments.

## 3 Artificial Intelligence and the Internet-of-Things (IoT)

The merging of Artificial Intelligence (AI) with Internet of Things (IoT) technology brings about a new era in cyber risk. This is marked by a complex interweaving of sophisticated threats that require an equally advanced approach to manage and mitigate them. This chapter delves into specific, technologically advanced examples that highlight the unique cyber risks brought about by AI in IoT environments, drawing from the foundational concepts in "Cyber Risk in IoT Systems."

One of the challenges posed by the use of AI in IoT is autonomous decision-making, which can amplify cyber risk. For example, AI-driven IoT devices in smart cities could autonomously manage traffic flow based on real-time data. However, a compromised AI algorithm could create chaotic traffic patterns, causing widespread disruption.

Data integrity is vital in AI-IoT systems, and data manipulation poses a risk. For instance, in healthcare IoT devices, AI algorithms process patient data for predictive diagnostics. The AI's predictive outcomes could be dangerously inaccurate if these data streams are manipulated – say through a man-in-the-middle attack intercepting and altering data from IoT health monitors. Similarly, AI model poisoning, where the AI's learning inputs are subtly tainted, could lead to erroneous learning, echoing the data integrity and manipulation concerns highlighted in the article.

Integrating AI into IoT brings unique AI-specific risks, such as adversarial machine learning. For example, in a network of interconnected smart home devices, an adversary could manipulate input data to an AI-powered security camera, causing it to misidentify or overlook intrusions. These AI-specific threats necessitate a novel approach to cybersecurity, diverging from traditional risk management strategies.

Addressing these enhanced risks requires a multifaceted and advanced approach. There is a need for risk assessment frameworks that specifically account for AI components in IoT ecosystems. This would involve understanding not only physical and data flow dependencies but also the AI algorithmic dependencies. Leveraging AI's capabilities for security in IoT networks presents a proactive defence mechanism. However, the implementation of such AI-driven security measures must be carefully managed to ensure they do not introduce new vulnerabilities. The integration of AI into IoT amplifies the need for comprehensive regulatory and ethical frameworks, addressing not only data privacy and security concerns but also the ethical implications of AI decisions, particularly in areas where these decisions impact human safety.

Given the complexity of AI in IoT, collaboration across disciplines is essential. Cybersecurity experts, AI researchers, IoT developers, and policymakers must work together to create advanced and resilient cybersecurity solutions that address the unique challenges posed by the AI-IoT convergence. In conclusion, the combination of AI and IoT presents a complex array of cyber risks that require advanced, specific, and comprehensive management strategies. Future research and practical approaches should focus on developing sophisticated AI-resilient security frameworks, enhancing regulatory standards, and promoting interdisciplinary collaborations, thus ensuring the secure advancement of AI within IoT systems.

## 4 Cyber risk from distributed ownership and control of IoT systems

The distributed ownership and control of IoT systems is considered the one factor contributing to the number of zero-day exploits exacerbated by IoT (Meakins 2019). Although there are many different cybersecurity approaches, they seem insufficient or not targeted at the right areas. This leads to a lack of security that creates unnecessary difficulty for IoT-connected producers and customers. The growth of the IoT market could increase significantly if policymakers have the methodology to assess, predict, analyse, and address the risks of IoT-related cyber-attacks in the communications network.

Without the appropriate risk assessment methodology, cyber risk could have costly consequences. Connecting cyber risk with IoT through impact models can provide feedback sensors and real-time data mechanisms to assist and enable industries and policymakers to understand and visualise the problem and address the risk created by IoT-related cyber-attacks.

### *4.1 Defining Cyber Risk*

IoT risk and the risk of cyber-attacks can be explained by established methods for calculating risk. Risk = Likelihood × Consequences, and cyber-risk can be defined as a function of:

R = { si, pi, xi }, i = 1, 2, ... , N,

Where R – risk; s – the description of a scenario (undesirable event); p – the probability of a scenario; x – the measure of consequences or damage caused by a scenario; N – the number of possible scenarios that may cause damage to a system.

The model above for calculating risk is classical (USA 2017), but the question remains how IoT risk and cyber-attack risk can be estimated. Since we do not have the precise measurements and concrete number of IoT cyber risks, an answer is difficult to present and justify with a desirable degree of certainty that the estimation is correct. Therefore, we discuss how IoT risk and the risk of cyber-attacks can be estimated assuming possession of the required data.

Businesses face strategic, compliance, operational, financial, and reputational risks regularly, all of which could affect their profitability or ability to function. Many businesses are looking to adopt new forms of technology (such as IoT, Blockchain and Artificial Intelligence) to increase the efficiency and effectiveness of their services. This exposes them to the risks that accompany these technologies.

While these technologies have the potential to improve their productivity, there is also the potential for the business to become increasingly susceptible to a series of security risks – this is the aspect of focus in this paper.

In the following table, we explore the main cyber risks that many businesses face, and we define definitions for different types of cyber risks. We use the term *'cyber risk'* in line with the Institute for Risk Management definition of: *'Cyber risk' means any risk of financial loss, disruption or damage to the reputation of an organisation from some sort of failure of its information technology systems.'* (Institute of Risk Management 2019). In Table 1, we provide insights on how companies can manage the IoT cyber risk, and we include real world examples in each risk category.

TABLE 1: DEFINING THE TYPES OF CYBER RISK FROM IOT SYSTEMS

| Types of Cyber Risk | Definition | Key Words | Example |
|---|---|---|---|
| Ethical | An action that falls short of what is considered morally right or outside of a professional standard. So, an ethical risk for an institution is the unintended harm caused by an unethical action. | Integrity, honesty, fraud, moral, standards, compliance, misconduct | Volkswagen develop and install software to cheat diesel emissions tests. This compromises organisation and industry standards and results in massive reputational and financial losses (Fracarolli Nunes and Lee Park 2016). |
| Privacy | Most information about people is digitised. Keeping this private and confidential is important. So, a privacy risk is when there is a temporary or permanent loss of control over data that may causes some form of harm to the individual and the business, organisation or government that holds the data. | Trust, transparency, data, confidentiality, cyberattack, hacking, data breach, phishing, pharming, ransomware, spam | In May 2014 145 million eBay customers have their names, addresses, date of birth and passwords accessed (Ranganthan et al. 2018).<br>Yahoo was the victim of a massive data breach in 2013-2014. In 2017 they finally admitted that all 3 billion user accounts had been compromised (Gupta 2018). |
| Security | This is to do with vulnerabilities and gaps in security programmes and systems. These vulnerabilities can be exploited in order to gain | Vulnerability, weakness, protection, attacks, breaches, DDoS (distributed | In October 2016 the Mirai Botnet launched a DDoS attack on DYN which led to parts of the internet going down and affected Twitter, Netflix, CNN, |
| | access to assets causing damage, harm or loss. Types of attack include physical, network, software and encryption attacks. | denial of service), botnets, malware, virus | Reddit (along with many others) (Dubois 2018; Payton 2018). Building management systems were attacked in a Finnish town of Lappeenranta, causing the heating in two buildings to fail (Scott and Spaniel 2016). |
| Technical | The failure of hardware or software due to poor design, construction, or evaluation. | Compliance, regulation, testing, evaluation | It was recently discovered that computer chips produced in the last 20 years all contain fundamental security flaws (Conte et al. 2018), some related to the chip variation, e.g., Spectre and Meltdown. |
| Interoperability | Refers to the challenges in ensuring that IoT devices, software platforms, or services can communicate effectively with each other. Interoperability risks can lead to system miscommunications or data integration failures, especially in environments with heterogeneous devices and platforms. | Compatibility, integration, communication breakdown, data silos | A healthcare system relying on IoT medical devices that operate on different protocols fails to integrate data from wearable sensors and in-hospital monitors, leading to incomplete or inaccurate patient health records. |
| Safety | Risks that directly affect the physical well-being of individuals or groups due to IoT device misuse, malfunction, or cyber-attacks. These risks include bodily harm or fatalities resulting from compromised safety-critical systems. | Physical harm, injury, malfunction, cyber-physical systems, personal safety | A compromised autonomous vehicle causes accidents due to failure in its IoT control systems, leading to injuries and potential fatalities. Similarly, hacked smart home devices such as IoT-connected thermostats cause fires or unsafe temperature regulation. |

Table 1 summarises and explores the ethical, privacy, security, and technical aspects of cyber risk. To relate the findings, an IoT-based example of this is the probability of a phishing attack on a connected corporate device occurring, like a company laptop or a smart phone, which then leads to the infection of that device. This infection could propagate to other IoT sensors in the company and consequently cause the disruption of their manufacturing plant's production line. While there are many application domains for IoT, for organisations to consider cyber security risk solely in the context of their domain would give misinformed results, since IoT is an ecosystem with platforms and services shared by different application domains.

### *4.2 IoT risk assessment*

Understanding what is meant by risk is only the first step when we are considering the potential risks in IoT. The next step is to be able to assess the risk, which involves the tasks of: (1) identifying (or defining) the risk - the action of developing a clear understanding of what organisational IoT assets are targeted by which threats and what harm could happen if those attacks are successful (Tanczer et al. 2018).

(2) Estimating the risk - this task aims to measure IoT risk based on the likelihood of the threat occurring and the impact on the organisation's infrastructure if it does occur. These measures can be qualitative (e.g., ratings using the levels, high, medium, and low) or quantitative (e.g., based on mathematical estimations and calculations).

(3) Prioritising the risk; once we have a list of the risks and each one has been estimated, the next task is for a company to prioritise the risks. This essentially provides a ranking of the risks based on their estimated levels. We interpreted that identifying the risk, estimating the risk, and prioritising the risk are three tasks of IoT risk assessment. **FIGURE 1**, below, sets this out, and demonstrates that this is a continuous process.

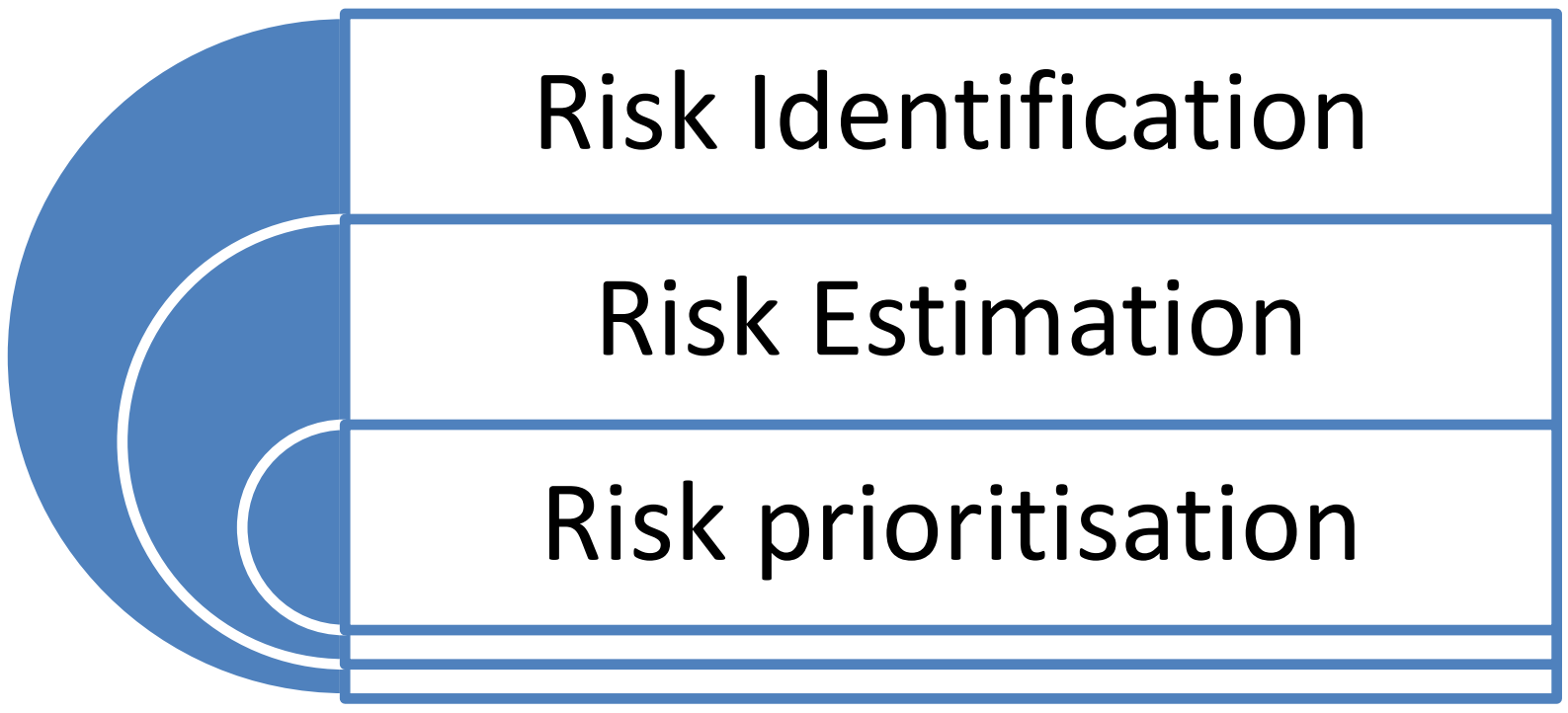

**FIGURE 1**. The IoT risk assessment process

### *4.3 IoT risk management*

The risk assessment process described above is part of risk management. While risk management techniques are well developed and used in various IT areas, there remains a significant challenge in managing IoT risk. Here, we include our findings in the form of four basic ways to resolve IoT risk:

IoT risk mitigation involves either reducing the likelihood of the risk happening or reducing the impact of the risk. In IoT risk management, this might include implementing IoT risk controls.

IoT risk transfer – this involves outsourcing the risk to a third party. In this instance, via cyber insurance for example;

IoT risk avoidance – this involves removing the risk. An example would be to remove IoT asset where the risk has originated; and

IoT risk acceptance – this involves accepting the risk as it stands, due to either the risk falling within the organisational risk appetite or the aggregated risk being sufficiently within the accepted risk levels.

The type of treatment selected for each risk is based on its estimated level, the costs associated with the treatment, and the organisation's overall tolerance for risk. In IoT, these factors are constantly changing, and this aspect represents one of the unique challenges when managing risks in dynamic IoT environments.

## 5 How IoT transforms the nature of risk

IoT represents interconnected technologies continuously communicating and sharing data. This technology creates serious safety risks and ethical concerns. For example, IoT incorporated into autonomous vehicles introduces safety risk, however, the device owner and the data owner are not necessarily the same (Anthonysamy, Rashid, and Chitchyan 2017), because there is no legal basis to actually own data. The data owner is the data curator or controller. Here we are making the point about the legal impossibility to own data. Because there is no owner of data, but rather an entity that has the legal right to control and steward the data. In following sections, we discuss how the existing risk assessment approaches can be adapted to assess the nature of IoT risk.

These designs need data to support, and the data is very sensitive and private. There has been a number of suggestions on how to resolve this concern. Back in 2014, the original 'Cyber Supply Chain Management and Transparency Act of 2014' (Royce 2014) was proposed and suggested that that US government agencies obtain a software bill of materials' (SBOMs) for all new software. This led to the 'Internet of Things Cybersecurity Improvement Act of 2017' (Howard 2017), and more recently, 'The US Executive Order on Improving the Nation's Cybersecurity of May 12, 2021, (Biden 2021) ordered The National Institute of Standards and Technology (NIST) to issue guidance on '*providing a purchaser a Software Bill of Materials (SBOM) for each product.'* These efforts in the US are related to resolving the specific issue of sharing sensitive and private company data on cyber vulnerabilities, exploits, threats, and this has been a very sensitive topic for a long time. The most recent effort that we are making to resolve these issues is the new the **Vulnerability Exploitability eXchange** (VEX)(NTIA 2021), which has already been adopted as a profile in the Common Security Advisory Framework (CSAF) (OASIS 2022). This article however, is more closely related to the updated version of the Common Vulnerability Scoring System Calculator (CVSS) (NIST 2022), which is the Stakeholder-Specific Vulnerability Categorization (CISA-SSVC) (CISA 2022) and it relates to the SSVC decision threes.

### *5.1 Security risk assessment for IoT systems*

One of the main problems with IoT is that this technology is developing at a fast rate and in multiple directions so that governments and national and international institutions face difficulties to standardize and enforce regulations in this field. These difficulties are related, for example, with the continuing changing environment of IoT (I. Brass et al. 2018) or with the relatively much slower legislative and standardization processes (e.g. (Schindler et al. 2013; Irina Brass et al. 2019)). We found that there are currently no risk assessment standards to govern companies in assessing the new types of risk before implementing IoT technologies and solutions. In the present climate, given the lack of unified global standards and regulations, businesses are pursuing economic profits from IoT solutions, but as it pertains to understanding the risk to their operations, businesses are often lacking in their approach to security.

### *5.2 Analysis of cyber risk assessment approaches*

As part of our research, we conducted an analysis of the existing cyber risk assessment approaches to enable us to provide basic guidance on how to develop a unified approach to risk assessment. Most cyber risk assessment approaches represent some similarities and after reviewing one we tend to get the general feeling that they all seem familiar. Hence, for differentiating these frameworks, for the reader and for our own research, in the Table 2 we tried to define the main differences between the cyber risk assessment frameworks that we reviewed in this article. In Table 2, we also include references to all of the frameworks as a source for further information on these frameworks. The selection process involved firstly conducting a literature review on the topic of 'most used' and 'most prominent' cyber risk assessment approaches.

***Secondary Data: 'most used' and 'most prominent'***

The selection of the 'most used' and 'most prominent' cyber risk assessment frameworks for this study was based on a combination of several key criteria that ensured relevance, industry adoption, and scholarly significance. These criteria were established following a comprehensive literature review and consultation with experts in cybersecurity, including those from Cisco Systems. The following criteria guided our framework selection:

1. **Industry Adoption and Standardisation**: One of the primary indicators of prominence was the degree of adoption within industry sectors and standardisation by international bodies. Frameworks such as **NIST Cybersecurity Framework** and **ISO/IEC 27001** were included because they are widely recognised and applied across various industries and sectors as global standards for cybersecurity risk management. Their extensive use across governmental, industrial, and private sectors made them foundational to this study.
2. **Scholarly Citations and Academic Relevance**: Frameworks that have been heavily cited in academic research and peer-reviewed journals were also prioritised. For example, frameworks like **OCTAVE** and **FAIR** have been the focus of numerous scholarly articles, making them prominent in the research community. The high citation count, particularly in the context of IoT cybersecurity and risk assessment, reinforced their relevance to this study's objectives.

3. **Expert Recommendations**: Insights from cybersecurity professionals and experts consulted during the research process, particularly those from Cisco Systems, played a crucial role in identifying frameworks that are "most used" in practice. These experts, with hands-on experience in cyber risk management, highlighted which frameworks they relied on in real-world scenarios, giving us a practical understanding of which frameworks are most relevant and widely applied across various sectors.
4. **Diversity of Application**: Frameworks that demonstrated applicability across a wide range of environments, including traditional IT infrastructures, IoT systems, and cloud computing, were considered more prominent. Frameworks such as **FAIR** (Factor Analysis of Information Risk) and **CVSS** (Common Vulnerability Scoring System) were selected because they are adaptable to different risk environments, including both qualitative and quantitative risk assessment contexts.
5. **Ease of Use and Implementation**: In practice, the complexity of a framework can influence its adoption. Frameworks that are well-documented, easy to use, and backed by automated tools or platforms were considered more prominent. For instance, **CVSS**, which provides a widely accessible scoring system for vulnerabilities, and **RiskLens**, which integrates FAIR for quantitative risk analysis, were selected for their ease of implementation in enterprise and IoT environments.
6. **Comprehensive Risk Coverage**: Finally, frameworks that cover a broad spectrum of risk factors, including technical, operational, strategic, and reputational risks, were included. The **NIST Cybersecurity Framework**, for example, is notable for its comprehensive approach, addressing everything from threat identification to incident response, which aligns with the holistic perspective of this study on IoT cyber risks.

The selection of frameworks was based on a multifaceted approach combining:

- **Industry recognition and standardisation**
- **Scholarly citation and academic significance**
- **Expert recommendations from cybersecurity practitioners**
- **Diversity and applicability across environments**
- **Ease of implementation**
- **Comprehensive risk coverage**

This selection process ensured that the frameworks chosen for analysis and inclusion in the study were not only theoretically sound but also practically relevant and widely used in the real world.

***Primary data: Expert consultations***

Secondly, we consulted a number of experts in the field from Cisco Systems that are responsible for this function. This consultation was conducted in the period between year 2018 and 2023, initiated with a scoping workshop in June 2018 and concluded with a closing workshop in January 2023. The consultation was conducted as case study action research, and included personal interviews with 43 cybersecurity experts, 13 workshops, two demonstration projects for gathering feedback, and 6 months long action research at Cisco locations.

The resulting list of approaches is not complete, but its representative of the 'most used' and 'most prominent' cyber risk assessment frameworks, models, and methodologies – as determined in literature and by the experts from Cisco Systems.

The Cisco Systems experts consulted during this study represented a broad range of cybersecurity specialities, including, but not limited to, cyber risk assessment frameworks. Their involvement was crucial in providing a comprehensive and multi-faceted view of IoT cyber risks and the development of robust risk assessment methodologies.

Specifically, a subset of the consulted experts specialised directly in **cyber risk assessment**, focusing on frameworks such as CVSS, CoSAI, OCTAVE, FAIR, and NIST Cybersecurity Framework, which were critical for refining the dependency model presented in this study. These experts were responsible for implementing and managing cyber risk strategies within Cisco's cybersecurity operations, making their insights particularly valuable in aligning the proposed model with industry practices and standards.

Additionally, the consultation involved professionals with expertise in **IoT security, network infrastructure vulnerabilities, and incident response frameworks**. Their contributions ensured that the proposed model incorporated a holistic understanding of the various layers of IoT ecosystems, including the unique challenges posed by real-time data flows, network management, and preventing cascading failures in IoT systems.

By engaging with a diverse group of experts, the study benefited from a broad spectrum of knowledge across different cybersecurity domains, ensuring that the proposed model focused on risk assessment and addressed practical implementation concerns, such as real-time threat detection, system recovery, and mitigation strategies. This interdisciplinary consultation strengthened the model's applicability to real-world IoT environments and enhanced its generalisation to diverse risk scenarios.

TABLE 2: ANALYSIS OF CYBER RISK ASSESSMENT APPROACHES

| Name | References to author(s) or Institution | What is it | Type |
|---|---|---|---|
| OCTAVE | (Caralli et al. 2007) | This is a standardised questionnaire that can be applied to investigate and categorise recovery impact areas. However, the OCTAVE method is complex and takes time to understand | Qualitative |
| TARA | (Wynn et al. 2011) | This is a predictive framework that enables targeting of the most crucial exposures, as opposed to promoting the defence of all possible vulnerabilities | Qualitative |
| CVSS (Common Vulnerability Scoring System) | (CVSS 2019) | A scoring system "that provides a way to capture the principal characteristics of a vulnerability and produce a numerical score reflecting its severity". It is relatively easy to use and translate results although, the calculator is based on experts' opinion and do not represent an ultimate precision, the calculator represents a guiding point. | Qualitative |
| Exostar System | (Shaw et al. 2017) | This system enables enterprises to assess, measure, and mitigate risk across multi-tier partner and supplier networks and determines the | Qualitative |
| | | gaps between cybersecurity posture and regulatory compliance. | |
| Capability Maturity Model Integration (CMMI) | (CMMI 2017) | This combines a set of best practices in the disciplines of systems analysis and design, software engineering and management. CMMI can simultaneously address multiple as opposed to stand alone improvements. This enables improvement in the entire enterprise risk and the full product development life cycle risk. | Qualitative |
| NIST Cybersecurity Framework | (NIST 2014) | This is a framework based on an extensive use of acronyms, which can be confusing and require a detailed understanding of the standards referred to in the acronyms. At present, the NIST framework is documented, not an automated tool. | Qualitative |
| ISO/IEC 27001 | (ISO 2017) | This risk management framework promotes standardisation of cyber risk and reflects international experience and knowledge. It is based on voluntary shared knowledge and is consensus based. | Qualitative |
| RiskLens | (FAIR 2020) | This is a quantitative assessment method based on FAIR (Factor Analysis of Information Risk) and "provides a model for understanding, analysing and quantifying information risk in financial terms". | Quantitative |
| CyVaR (Cyber Value at Risk) | (Cyberpoint LLC, n.d.) | This presents a method to quantitatively assess risk with Monte Carlo simulations. CyVaR needs to be adapted and modified to include units of measurement for IoT cyber risk vectors. | Quantitative |
| FAIR | (FAIR 2017) | This model promotes a quantitative, risk based, acceptable level of loss exposure. | Quantitative |

The analysis in table 2 provides guidance and concludes that most of the cyber security frameworks today apply qualitative approaches to measuring cyber risk, while quantitative approaches are mostly present in the cyber security models. The analysis in table 2 also confirms that none of these approaches resolves adequately the cyber risk assessment in IoT, at least not individually or in isolation. Presented with the diversity of cyber risk assessment approaches analysed in Table 2 and given that existing risk methods do not address entirely the cyber risk from IoT, questions emerge on: a). how can these approaches be combined into a unified model, and b). how can we be certain that a unified model addresses IoT context. We try to address these questions in section 4.2 through a dependency model that presents a unified approach for improved standards, governance, and policy on data strategies.

## 6 Dependency Modelling for creating a unified model

In this section, a unified cyber risk assessment approach for IoT risk is explored via dependency modelling (DM) approach and a step-by-step process is included, enabling other companies to replicate this cyber risk assessment process. Dependency modelling (DM) is a goal-oriented method of representing the interactions and inter-reliance amongst system components or elements using same to reason about the scope of risk feasible (Cherdantseva et al. 2022). DM works on the assumption that risks emerge from interactions and interdependencies which need to be recognised in order to effectively manage and guard against the impacts of the risks (Alpcan and Bambos 2009). DM for security risk assessment can work through analysing the vulnerabilities that can be found in IoT network/system components – evaluating the interactions and service flow amongst connected components including hardware infrastructure, software platforms (applications), processes, services, users, etc., and how these threats and and vulnerabilities affect both the target components and others connected. Generally, these are explored considering how the entire system functions and objectives are impacted. Security threats and vulnerabilities can emerge or exist in diverse forms, ranging from design flaws in hardware, software, and processes, as well as competency limitation in users, which can easily be exploited by malware, social engineering, etc. Thus, the service or functional dependencies amongst IoT system components can be used to design a unified approach for IoT risk assessment. In doing this, we consider contexts from IoT literature and use cases in the model definition and verification.

Interactions within IoT can be seen as complex, tightly coupled relationship structures amongst the systems, sub-systems, and components. This means that IoT subsystems and components inter-cooperate to fulfil desired service objectives, which each sub-system or components is unable to achieve in isolation. To function appropriately, one or more sub-systems rely considerably on the appropriate functions of another system or sub-system they connect to and receive command or instruction input. Thus, a dependency relationship (shown in **FIGURE 2**) exists between connected systems with a mechanism characterised by the transfer of data or control from one component to another (Callo Arias, Van Der Spek, and Avgeriou 2011), and which can either be direct (a first order) or indirect (a higher order) (Laugé, Hernantes, and Sarriegi 2015), physical or non-physical (O'Neill 2013), and involve any constituent of the wider IoT System operational ecosystem.

Graph theoretical approach can be used to represent dependencies in IoT networks as shown in **FIGURE 2**. This presents a directed graph structure G as an ordered pair (C, T), where C represents a finite set of vertices referring to IoT components, T representing a binary relation on C. T imply edges which represent *'context transfer or flow'* along successive IoT components. These edges form an ordered pair t = ($c_i$, $c_k$), where $c_i, c_k \in C$ represent interacting or cooperating IoT components on specific functional objective. t can represent the dependency flow of data, service, or functionality from an originating IoT component $c_i$ to a destination IoT component $c_k$ (See **FIGURE 2**).

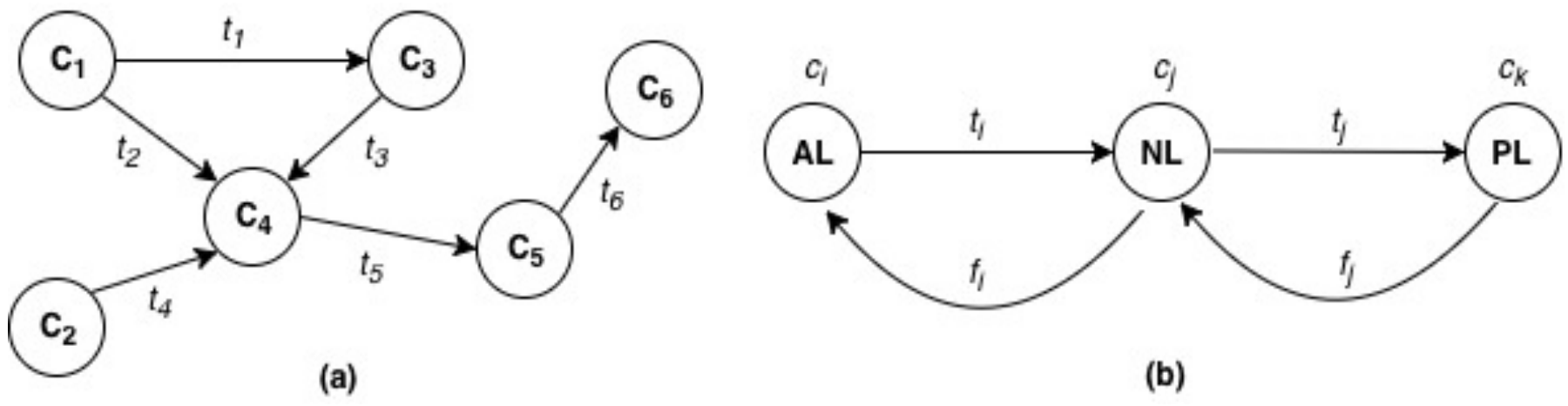

**FIGURE 2.** Dependency relationships are indicated by the dependency expressed in the directional arrows.

This dependency relationship could apply to different types of cyber risks. Since our efforts are focused on different types of risk assessment, including cyber risk assessment in general, we have used examples specific to IoT risk. Firstly, take, for example, an industrial Internet-of-Things (IIoT) production line involving a robotic arm and a conveyor belt system for product identification, transfer, and packing, following the analogy. To optimise packing performance, desired analytics functions by cloud-based components and services such as HMI and performance dashboard on the application layer (AL), which can represent $c_i$ of C in an IIoT system, would typically depend on the appropriate functioning of transmitted data $t_i, t_j \in T$ through the network layer (NL) components such as communication switches and Programmable Logic Controllers ($c_j$). Dependency could also extend to perception layer (PL) components such as the Photoelectric sensors that detect items and actuator switches that move conveyor belts ($c_k$).

Like in other digital systems, IoT security risks typically depend on the existence/exploitation of vulnerabilities in system components at any layers of the architecture. Exploiting vulnerable components can cause them to malfunction or fail to deliver the desired processes initially configured. If an attacker gains control of sensing and/or actuating service functions and flow $t_j$ on a PL, wrong data could be transmitted to and through N, and worse, data flow could be completely stopped. The impact on process data can reach AL components such as HMI and performance analytics dashboard system, and can in turn impact on the functions or outputs ($t_i$) desired from components in AL. The impacts can include a failure to reach the final goal of passing down correct item processing data for analysis and optimisation functions to support decision-making. If a vulnerability on a host IoT component is exploited, potential functional dependency-based impact can be estimated quantitative as a proportion of an overall flow of component functional dependencies along the part of compromise. A functional dependency index can be evaluated by analysing the number of components that are included along a path following the edges from the originating component. Depending on the existence or otherwise of a functional dependency link, initial impact(s) of the attack is typically expressed in the origin and flows through to other connected components along the same path.

A logical switch function $\varphi(v)$ can be used to evaluate the conditional existence of a functional dependency between any two nodes on the network, a logical 0 (FALSE) to indicate ‘connection not configured’, and a 1 (TRUE) to indicated connection configured as shown in Equation 1. For a tree network structure for IoT, the functional dependency index $fd_v$ of a component v can be evaluated by summing the functional dependency indices of components connected to component v with a ‘connection configured’ settings, as shown in Equation 2.

| | | |
|---|---|---|
| $\varphi(v) = \begin{cases} 1 \\ 0 \end{cases}$ | $\rightarrow$ *connection configured*<br>$\rightarrow$ *connection not configured* | *(1)* |

| | |
|---|---|
| $$fd_v = \sum_{u \epsilon T_v} (fd_u \times \varphi(v))$$ | *(2)* |
| *where, $T_v$ represents the subset of components reachable directly from v.* | |

**FIGURE 3**. Connection configured

The proportion of impact dependency can be evaluated in relations to the highest possible dependency, which represents worse case impact of a vulnerability exploitation max $(fd_v)$. A worst-case scenario can involve a dependency that runs through all the components in IoT network, enabling negative impacts to also flow along the same path when a certain vulnerability is exploited. Here, the impact dependency proportion would be 1. A 0 would mean no component is affected.  An impact dependency proportion, $(P_{fd_v})$ can be estimated as the degree of dependency impact which can occur when a certain vulnerability is exploited relative to the worst-case dependency impact (See Equation 3).

| | |
|---|---|
| $$P_{fd_v} = \frac{fd_v}{max(fd_v)}$$ | *(3)* |

**FIGURE 4**. Worst-case dependency impact

Thus, an IoT security risk landscape need not consider the failure of a single IoT component alone, but the failure of other IoT components (devices or services) due to abnormal events and/or impacts on a component they rely on. Functional dependency relationships amongst IoT sub-systems can also cause impacts or failures to cascade from one affected system or component onto another; aggravating the impacts (Bloomfield et al. 2010).

Depending on the evaluation approach, security and safety-critical impacts typically vary amongst assets, their functionalities (services), placement positions, and configurations within industrial networked systems, including IoT. However, to support effective decision-making from both security and safety perspectives, IoT adopters need to adopt risk assessment methods that goes beyond considering vulnerability/risk scenarios one-by-one, qualitatively or statically, to considering the relationship between the risk factors. This can provide a more thoughtful understanding of the scale of impacts involved and drive appropriate prioritisation of security controls and responses. The dependency relationships are indicated by the directional arrows (in **FIGURE 2**), where the expressed dependencies describe a model for addressing IoT security risks.

The process of measuring the probability of things breaking down or dependencies is well understood in cyber economics, and many papers have made an effort to calculate these numbers and provide ROI. Although some would argue that they are limited, but the evidence of such publication confirms that the lack of probabilistic data has not stopped either firms or researchers to make an effort. Hence, in this paper we try to relate similar efforts towards the assessment of IoT risk, by repeating similar thoughts throughout the paper.

To be impactful, risk assessment method needs to consider intrinsic capabilities as well as the more general characteristics of the IoT system which enable security risks. Capabilities can range from sensing, processing, actuating, interfacing, storage, and usage management. Characteristics can range from component heterogeneity, scale variability, connection temporality, low power retention, and intelligence generation/fusion.

This way, they can achieve the quantification of security-related dependencies that can help provide deeper and better security insights. Some of these insights include understanding how the impact of exploiting certain security vulnerability(ies) in an IoT infrastructure component or subsystem prevents it from delivering the relevant and required service(s), and how such affects the performance of other connected sub-systems that connect to, require data/service flows from, and rely on an affected target.

This can help in the development and adoption of effective security incident response and recovery (Laugé, Hernantes, and Sarriegi 2015), as well as help reduce and manage the effects of IoT disruptions.

## 7 Cyber risk acceptance and transference – response and recovery

The argument for using the dependency model to assess risk in IoT sub-systems is that we can also assess the impacts caused by failures that cascade through the system and understand the scale of such impact in relation to the fulfilment of operation objectives. Depending on the outcome of evaluating functional dependencies, after all possible states have been considered, often, there is a possibility of a ‘no-win’ incidents, where each scenario leads to a risk that cannot be totally controlled or eliminated. The next states would be risk acceptance and risk transference. In the following section, we give an overview of the steps involved in risk acceptance - which includes incident response, recovery (Van Kleek et al. 2018), and we end this section with a discussion of cyber insurance, which represents a method for risk transference.

### *7.1 Risk Acceptance - Incident Response and Recovery*

In this section, we provide insights on how companies can manage risk through their incident response and recovery. One form to describe risk acceptance is a state where adding additional defences becomes too expensive within a certain dependency model. In such a state, it could be rational to accept that future attacks will happen. Then, an initial zero defence configuration is supported with a reactive defence that is activated when a vulnerability is exploited (Woods and Simpson, n.d.). Incident response and recovery for IoT can follow similar approaches already common to digital and computing systems. Key phases of an incident response and recovery procedure for IoT systems include planning, detection, analysis and response formulation, containment, eradication, recovery, and post-incident activity. The diagram in **FIGURE 5** (below) illustrates this process.

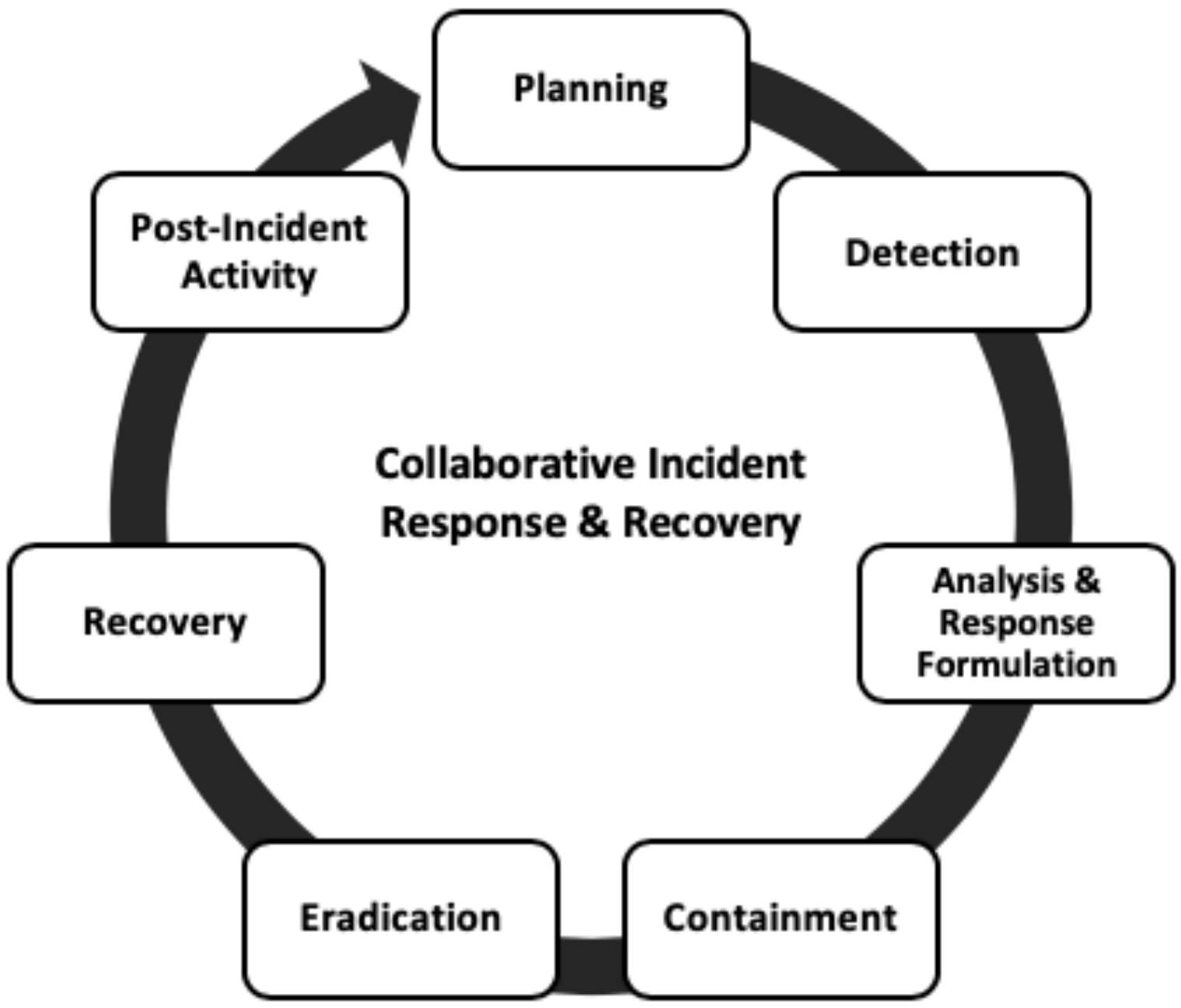

**FIGURE 5**. The new emerging security Incident Response Approach

Companies can use the Planning phase or Incident Response (IR) Preparation, which involves activities that ensure IoT-user organisations are in a state of readiness for the prompt handling of incidents. Example activities include:

- Setting and training IoT incident response teams;
- Defining appropriate security response and escalation policies;
- Forensic evidence management in line with relevant security guidelines and practices;
- Awareness and response strategies to common security threats such as Denial of Service, Worms/Trojans, Phishing, web-based and web application attacks, insider threats, exploit kits, information leakage, and identity theft.

However, the scope and number of questions that need consideration and answers have only increased in planning for an effective response to IoT security incidents compared to traditional IT systems. For example, a self-driving vehicle relies on radar sensors to detect obstacles, which evidently can fail (Breza, Tomic, and McCann 2018), resulting in a crash. This is a complex human-machine system relying on many different systems owned by a variety of enterprises. A good solution path would include viewing IoT incident response considering the mission process of IoT devices and system. This can support understanding of how malicious actors might exploit the normal infrastructure (device or system) functionality failures and impacts to hide malicious actions. Greater understanding can be achieved through ensuring that security experts responsible for securing the operations of IoT systems clearly understand the threat models that drive the systems (Russell and Van Duren 2016). Security experts also need to be conscious and responsive to the dynamic states of the security threats to provide effective response actions.

Several conventional IR methodologies and frameworks can be adapted for IoT. While it may not be feasible to prevent all security compromises in IoT, effective threat response and management are needed. These must be built on well-structured and holistic incident response plans and procedures and on the respective dependency models assumed/ used.

Companies can also use the Detection phase of IR, which emphasises the importance of promptly recognising the beginning of what is considered a ‘threat’ in an IoT system for which critical decisions and actions are required.  Since IoT relies on cloud-hosted infrastructures and often includes limited-functionality devices (from an events and log management perspective), it is necessary to include in the infrastructure monitoring design a capacity to capture instrumentation data directly from IoT devices, as well as from supporting cloud service providers. IoT devices use trusted credentials for exchanges, which, when compromised, can result in significant impacts across the system. As described above, only complex monitoring can provide the visibility necessary to spur timely decisions and responses. There is an increasingly significant role for computational intelligence in supporting risk assessment through identifying risk, capitalising on opportunities, and gaining a deep understanding of a business through reports, dashboards, visualisations, and information analysis.

Traditional security information and event management (SIEM) systems, although powerful and well-advanced for standard networks, are unable to handle the complexities involved in IoT, where massive numbers of nodes and millions of data are involved. Hence, the need for newer, more tailored IoT-centric systems.

Instead, Companies can use the Analysis and Response formulation phase, which focuses on understanding the characteristics of security threats or incidents to learn the most suitable strategy or method for handling future incidents; again, traditional systems struggle, and IoT-specific digital forensic and incident response tools are necessary. In IoT, analysis of threats should consider both system-wide and component-specific perspectives. Using effective threat intelligence tools and processes that relate to IoT application sector is a good place to start, as threat indicators and protective patterns are often shared and made available on threat intelligence platforms.

From these, further analysis can be explored evaluating the scope of compromise, activities, timelines, and attacker identities related to certain breaches. However, recognition of the potential for attacks to employ anonymity and other anti-forensic capabilities characterised in the IoT domain is required. Since IoT systems are data-intensive, data compromise analysis with respect to confidentiality, integrity, and availability is also crucial. These mechanisms for assuring integrity and availability can be complemented with IoT devices and gateway forensic analysis to provide acceptable proof of the breach of IoT devices and systems.

Alternatively, companies can use the Containment phase, which aims to ensure prompt, interim resolution to a security incident by engaging in attempts to restrict further damage to the system. Typical actions in traditional IT systems may include disabling affected services, disconnecting or swapping out compromised devices and systems with new ones, revising access credential values such as passwords, disabling affected accounts or, at worst, initiating a temporary shut-down. Some of these activities do not translate to incidents in IoT systems; here we list them as descriptive examples. The main task in this phase is for affected devices, services, or systems to be isolated from the operations IoT network as quickly as possible while allowing for forensic analysis of affected systems.

Companies can also use the Eradication phase, which leads from the containment phase, focuses on the long-term removal of threats, and ensures that the system is no longer vulnerable to the threat. Typical activities in this phase include policy updates and independent security audits. This can be achieved in IoT systems by evaluating whether existing security policies can sufficiently address any threats that have been identified; if not, security policy upgrades need to be adopted and implemented. For example, automated software/firmware updates and patching are challenging in today’s IoT. It is necessary to devise and adopt policies and approaches for security patching that would provide the necessary security without disrupting operations and functionality. With reference to the need to support forensic analysis, it is desirable to track the activities of a malicious actor in a network. IoT can benefit from gateway devices that support the establishment of logical rules for automated isolation of compromised infrastructure based on monitored commands or traffic flow patterns without alerting an active attacker on the network (Craggs and Rashid 2017). In this way, the attacker's actions and activities can be observed and studied to inform decisions for necessary security improvements.

Companies can use the Recovery phase to restore the system to normal working order. Typical actions may include restoring systems using backups, system re-configurations, or fresh installations. These must be considered for both cloud and on-premises infrastructure, and restoration must be initiated in a way that does not cause significant delays or disruptions to the normal operation of the IoT system.

Companies can use the Post-Incident Activity phase, which includes a combined process of drawing lessons from breaches and reporting these lessons in a structured way that helps to form capability for future occurrences. Typically, this should be conducted through reflective meetings that bring together senior executives and technical experts (Falco, Noriega, and Susskind 2019). In the reflective reviews, privacy checks, root cause analysis, and after-incident forensics can be performed in relation to the compromised system. Using root cause analysis, organisations can easily understand the failure of their security and determine how to strengthen the weaknesses as well as produce true assessments of what happened, how it happened, how well or poor the response went and why, and what a better response may look like in the future. Overall, lessons learned should be evaluated and amended as required, including the incident response plan, the network access control (NAC) plan, existing tools and resources to enhance security, deficiencies in cloud service providers and the on-premises incident response process.

IoT brings inherent cyber risks spanning multiple functional sectors with varied dependencies. Further, IoT systems often operate on platforms that cut across geographical boundaries for which appropriate cyber incident response and recovery plans and strategies are required. Collaborative Incident Response and Recovery (IR&R) utilises shared threat intelligence and should evolve based on this intelligence. This is required since the security risk landscape is continually evolving, so an incident response plan which was appropriate yesterday might not be today; a plan that seems effective today could also be ineffective tomorrow. Effective IR&R should (1) be designed to fit the dependency model chosen to assess risk in the respective IoT environment or service, and (2) be characterised by continuous refinements of processes and procedures. This represents a move from a reactive response to the management of security incidents in a way that fosters cooperation through the exchange and sharing of incident management information among several distinct IoT-adopter organisations. Such an approach should enable both proactive and reactive capacities and enforce and assure trust and privacy among IoT infrastructures and cooperating organisations.

These findings represent a key insight that refers to a wide variety of enterprises, and it addresses a missing discussion of the impact of IoT cyber risk on liability and insurance risk ownership. The answer must be partially addressed by virtual reality cyber assessment (Furfaro et al. 2017) and cost and frequency analysis of cyber-attacks. Such analysis would complement building frameworks and methodologies for mitigating the impact of cyber risk and assessing cyber risk in IoT-connected products and services. This would resolve the previously discussed lack of standardised methods for measuring the cost and probabilities of cyber-attacks in IoT systems and the impact of such (IoT product, service or platform-related) cyber risk. The lack of empirical data to construct actuarial tables applies to cyber risk in general. Adding to this, the growth of IoT cyber risk markets in the finance and insurance sectors is impeded by the lack of empirical data to construct actuarial tables (Egan et al. 2019). We could also argue that actuarial tables are irrelevant in many emergent risk markets - for example, cyber insurance creates what is called 'reliance' - that is, reliance that insurance companies take care of possible risks or financial risk depends not necessarily on actuarial tables, but rather on specific mechanisms such as how the markets price the potential hazards and price the consequences.

Nevertheless, the highly dynamic systems in these sectors make it difficult for businesses to formulate significant assumptions on the nature of risk, as even the possible knowledge of risks can further affect them. Despite the development of models related to the impact of cyber risk (Jalali et al. 2019; Evans 2019), there is a lack of such models related to specific IoT verticals. Hence, banks and insurers cannot price IoT cyber risk with the same precision as in traditional insurance lines (Camillo 2017).

## 8 Case study discussion on estimation and valuation of IoT cyber risk

While conducting this research, we used the case study and action research methods to apply our research findings in practice. Since this research was co-funded by Cisco Systems, one of the main benefits of this research was the access and engagement with their cyber risk management. We used their risk management tools as a platform to test, verify and advance our understanding of the role IoT is playing in their risk management operations. One of the first case study discoveries was related to risk transference and how companies are dealing with such unpredictable risks. Cyber risk insurance represents risk transference and is categorised as a risk management operation. IoT technologies are becoming more prevalent, and we can observe cyber risks worldwide, increasingly impacting physical property and challenging present notions of accountability and liability.

Consequently, cyber insurance has often been investigated as a possible market-based solution to cyber security problems. For example, in dynamic systems, cyber insurance is meant to control financial risk and thus depends on how the markets price the possible hazards and the consequences. However, the cyber insurance market needs help in measuring and assessing risks and designing and managing cyber risks efficiently. Some of the major problems cyber insurers face is the lack of historical data on risks, a lack of claims data, the volatility of the rather immature IoT technology and markets and the increased scope for cyber security risks. From a broader perspective, governments and the insurance industry are far from a working public-private partnership for cyber insurance.

- To identify how a company can deal with such risk scenarios, we conducted action research with Cisco Systems. From our case study research, we identified that our model for risk assessment could be applied if we had the probabilistic data we do not have. Therefore, in our action research, we focused on the data strategy. We have worked with Cisco Systems for three years and developed a data strategy to deliver the probabilistic data needed for risk assessment. This data strategy was presented to the FAIR Institute webinars, and we gathered further feedback from other companies. The advantage of participating with the FAIR Institute was that we gained access to many different companies' specific cyber risk departments. In Table 3, we include a snapshot of the simulation of the proposed goal-oriented approach. The original table is a much larger document, and the image we see in Table 3 is just a small sample to demonstrate the process. What we can see in the demonstration is a unique code for each risk category (on the left side), where each risk category is allocated to a specific principle, and principles are categorised in areas of focus. Individual principles are allocated weights from 0 to 3, and the weight is determined by the risk exploitability of the vulnerabilities allocated to the specific principle. Applying the design to the previously described goal-oriented approach is necessary, which also operates as a decision three in this scenario.

## TABLE 3: SIMULATION OF THE GOAL-ORIENTED APPROACH

| | Areas | Principles | SecurityCulture | | | | | CriticalInformation | | | DigitalGovernance | OperationalExcellence | | | | GroupInformation | TechnologyDefence | | | | DigitalWorkplace | DataProtection | | | Identity&AccessManagement | | | | | | | IncidentResponse | | | |
|---|---|---|---|---|---|---|---|---|---|---|---|---|---|---|---|---|---|---|---|---|---|---|---|---|---|---|---|---|---|---|---|---|---|---|---|
| | | | AdvancedSecurityTraining | E-learningsSecurityLab | AwarenessCampaign | Anti-Phishing | InsiderThreats | AudittrailsLogging | Penetrationtesting | Encryption | ComplianceGates | Regulations | SecureApplicationDevelopment DevSecOps | AssetManagement | Innovation | Riskassessmentsandreviews | AdvancedThreatAnalytics | ITServiceControl | SecureConfigurationandVulnerabilityScanning | Perimetervulnerabilityscanning &penetrationtestingWWPT | Windows10 | InformationClassification | DigitalRightsManagement | DataLeakagePrevention | PAMInfrastructureAdmin | PAMPrivilegedBusinessUser | ExternalUserManagement | IAMforCriticalInformation | Anomalydetection | Simplificationstrategy | 0trustenvironment | SimulationexercisesIMTDPMEMTGCMTBOD | RedPurpleTeaming | PreparednessGroupIRPlanincl. op.playbooks | CompromiseAssessment |
| #0097A29 | **SecurityCulture** | AdvancedSecurityTraining | | 0 | 0 | 0 | 0 | 3 | 1 | 2 | 3 | 3 | 1 | 2 | 3 | 1 | 3 | 1 | 2 | 3 | 2 | 2 | 3 | 1 | 1 | 2 | 3 | 1 | 2 | 3 | 1 | 1 | 2 | 3 | 1 |
| #0097A28 | | E-learningsSecurityLab | 0 | | 0 | 0 | 0 | 2 | 3 | 1 | 2 | 2 | 3 | 1 | 2 | 3 | 2 | 3 | 1 | 2 | 3 | 1 | 2 | 3 | 3 | 1 | 2 | 3 | 1 | 2 | 3 | 3 | 1 | 2 | 3 |
| #0097A27 | | AwarenessCampaign | 0 | 0 | | 0 | 0 | 3 | 1 | 2 | 3 | 3 | 1 | 2 | 3 | 1 | 3 | 1 | 2 | 3 | 1 | 2 | 3 | 1 | 1 | 2 | 3 | 1 | 2 | 3 | 1 | 1 | 2 | 3 | 1 |
| #0097A26 | | Anti-Phishing | 0 | 0 | 0 | | 0 | 2 | 3 | 1 | 2 | 2 | 3 | 1 | 2 | 3 | 2 | 3 | 1 | 2 | 3 | 1 | 2 | 3 | 3 | 1 | 2 | 3 | 1 | 2 | 3 | 3 | 1 | 2 | 3 |
| #0097A25 | | InsiderThreats | 0 | 0 | 0 | 0 | | 2 | 3 | 1 | 2 | 2 | 3 | 1 | 2 | 3 | 2 | 3 | 1 | 2 | 1 | 1 | 2 | 3 | 3 | 1 | 2 | 3 | 1 | 2 | 3 | 3 | 1 | 2 | 3 |
| #0097A24 | **CriticalInformation** | AudittrailsLogging | 3 | 3 | 3 | 3 | 3 | | 0 | 0 | 1 | 2 | 3 | 1 | 2 | 2 | 3 | 1 | 2 | 3 | 1 | 2 | 3 | 1 | 3 | 1 | 2 | 3 | 1 | 2 | 3 | 3 | 1 | 2 | 3 |
| #0097A23 | | Penetrationtesting | 1 | 3 | 3 | 3 | 3 | 0 | | 0 | 2 | 3 | 1 | 2 | 3 | 2 | 2 | 3 | 1 | 2 | 3 | 1 | 2 | 3 | 3 | 1 | 2 | 3 | 1 | 2 | 3 | 3 | 1 | 2 | 3 |
| #0097A22 | | Encryption | 3 | 3 | 3 | 3 | 3 | 0 | 0 | | 3 | 3 | 3 | 3 | 3 | 3 | 3 | 3 | 3 | 3 | 1 | 1 | 2 | 3 | 3 | 3 | 3 | 3 | 3 | 3 | 3 | 3 | 3 | 3 | 3 |
| #0097A21 | **DigitalGovernanceFramework** | ComplianceGates | 3 | 3 | 3 | 3 | 3 | 3 | 3 | 3 | | 3 | 3 | 3 | 3 | 3 | 3 | 3 | 3 | 3 | 3 | 3 | 3 | 3 | 3 | 3 | 3 | 3 | 3 | 3 | 3 | 3 | 3 | 3 | 3 |
| #0097A20 | **OperationalExcellence** | Regulations | 3 | 3 | 3 | 3 | 3 | 3 | 3 | 3 | 3 | | 0 | 0 | 0 | 3 | 3 | 3 | 3 | 3 | 3 | 3 | 3 | 3 | 3 | 3 | 3 | 3 | 3 | 3 | 3 | 3 | 3 | 3 | 3 |
| #0097A19 | | SecureApplicationDevelopı | 2 | 2 | 2 | 2 | 2 | 2 | 2 | 2 | 2 | 0 | | 0 | 0 | 2 | 3 | 1 | 2 | 3 | 1 | 2 | 3 | 1 | 3 | 1 | 2 | 3 | 1 | 2 | 3 | 1 | 2 | 3 | 1 |
| #0097A18 | | AssetManagement | 1 | 2 | 1 | 3 | 1 | 1 | 2 | 1 | 3 | 0 | 0 | | 0 | 3 | 2 | 3 | 1 | 2 | 3 | 1 | 2 | 3 | 1 | 2 | 3 | 1 | 2 | 3 | 1 | 3 | 1 | 2 | 3 |
| #0097A17 | | Innovation | 1 | 3 | 2 | 1 | 1 | 3 | 3 | 3 | 3 | 0 | 0 | 0 | | | 3 | 3 | 3 | 3 | 3 | 3 | 3 | 3 | 3 | 3 | 3 | 3 | 3 | 3 | 3 | 3 | 3 | 3 | 3 |
| #0097A16 | **GroupInformationSecurity** | Riskassessmentsandreview | 1 | 3 | 2 | 1 | 1 | 2 | 2 | 3 | 2 | 2 | 3 | 1 | 2 | | 2 | 3 | 1 | 2 | 3 | 1 | 2 | 3 | 1 | 2 | 3 | 1 | 2 | 3 | 1 | 1 | 2 | 3 | 1 |
| #0097A15 | **TechnologyDefence** | AdvancedThreatAnalytics | 1 | 3 | 2 | 1 | 1 | 3 | 3 | 3 | 3 | 2 | 3 | 1 | 2 | 2 | | 0 | 0 | 0 | 3 | 1 | 2 | 3 | 3 | 1 | 2 | 3 | 1 | 2 | 3 | 3 | 1 | 2 | 3 |
| #0097A14 | | ITServiceControl | 3 | 3 | 3 | 3 | 3 | 2 | 2 | 2 | 2 | 3 | 1 | 2 | 3 | 2 | 0 | | 0 | 0 | 1 | 2 | 3 | 1 | 3 | 3 | 3 | 3 | 3 | 3 | 3 | 3 | 1 | 2 | 3 |
| #0097A13 | | SecureConfigurationandVu | 1 | 3 | 3 | 3 | 3 | 3 | 3 | 3 | 3 | 2 | 3 | 1 | 2 | 1 | 0 | 0 | | 0 | 3 | 1 | 2 | 3 | 3 | 1 | 2 | 3 | 1 | 2 | 3 | 3 | 3 | 3 | 3 |
| #0097A12 | | Perimetervulnerabilityscan | 3 | 3 | 3 | 3 | 3 | 2 | 2 | 2 | 1 | 3 | 1 | 2 | 3 | 3 | 0 | 0 | 0 | | 3 | 1 | 2 | 3 | 1 | 2 | 3 | 1 | 2 | 3 | 1 | 1 | 2 | 3 | 1 |
| #0097A11 | **DigitalWorkplace** | Windows10 | 3 | 3 | 2 | 2 | 2 | 1 | 2 | 1 | 3 | 2 | 3 | 1 | 2 | 3 | 2 | 3 | 1 | 2 | | 2 | 3 | 1 | 3 | 1 | 2 | 3 | 1 | 2 | 3 | 3 | 3 | 3 | 3 |
| #0097A10 | **DataProtection** | InformationClassification | 3 | 3 | 3 | 3 | 3 | 2 | 2 | 3 | 2 | 2 | 3 | 1 | 2 | 3 | 2 | 3 | 1 | 2 | 3 | | 0 | 0 | 1 | 2 | 3 | 1 | 2 | 3 | 1 | 3 | 1 | 2 | 3 |
| #0097A9 | | DigitalRightsManagement | 1 | 3 | 3 | 3 | 3 | 3 | 3 | 3 | 3 | 2 | 3 | 1 | 2 | 2 | 3 | 1 | 2 | 3 | 2 | 0 | | 0 | 3 | 1 | 2 | 3 | 1 | 2 | 3 | 3 | 3 | 3 | 3 |
| #0097A8 | | DataLeakagePrevention | 3 | 3 | 3 | 3 | 3 | 2 | 2 | 2 | 3 | 3 | 1 | 2 | 3 | 3 | 2 | 3 | 1 | 2 | 1 | 0 | 0 | | 1 | 2 | 3 | 1 | 2 | 3 | 1 | 1 | 2 | 3 | 1 |
| #0097A7 | **Identity&AccessManagement** | PAMInfrastructureAdmin | 1 | 1 | 2 | 1 | 1 | 2 | 2 | 3 | 3 | 2 | 3 | 1 | 2 | 1 | 3 | 1 | 2 | 3 | 2 | 2 | 3 | 1 | | 0 | 0 | 0 | 0 | 0 | 0 | 1 | 2 | 3 | 1 |
| #0097A6 | | PAMPrivilegedBusinessUse | 2 | 2 | 1 | 3 | 1 | 3 | 3 | 3 | 2 | 3 | 1 | 2 | 3 | 3 | 2 | 3 | 1 | 2 | 3 | 1 | 2 | 3 | 0 | | 0 | 0 | 0 | 0 | 0 | 3 | 1 | 2 | 3 |

In Table 3, we can visualise the process of applying the proposed goal-oriented approach. The unique code is also a unique reference to a specific vulnerability that is found in the National Vulnerability Database (NVD), which are stored as JSON files. The unique code is included to resolve the product naming problem, which is one of the most difficult issues to solve in the new software bill of materials (SBOM) and the proposed integration with the vulnerability exploitability exchange (VEX). This work relates to the ongoing efforts of the Common Security Advisory Framework (CSAF) and the new Stakeholder-Specific Vulnerability Categorisation (SSVC), which is an updated version of the Common Vulnerability Scoring System Calculator (CVSS). Still, it's based on a decision threes and qualitative data.

**Advantages of SSVC's Decision Trees and Qualitative Data for Prioritising Vulnerabilities**

The Stakeholder-Specific Vulnerability Categorisation (SSVC) applies decision trees and qualitative data, and offers several key advantages for prioritising vulnerabilities in a goal-oriented approach to IoT cyber risk management. While traditional risk assessment methods often rely heavily on quantitative data, the inclusion of qualitative assessments through SSVC enhances flexibility, adaptability, and relevance to real-world IoT systems, where data may not always be complete or measurable in a purely quantitative form. Below are the primary advantages of SSVC in this context:

**1. Tailored Decision-Making Process**

One of the strengths of SSVC's reliance on decision trees is that it enables the prioritisation of vulnerabilities based on context-specific factors relevant to each organisation's risk tolerance and operational environment. The decision tree methodology provides clear decision points, such as whether a vulnerability needs immediate patching or whether it can be delayed based on factors like:

- The potential impact on critical services,
- The presence of mitigations, or
- The likelihood of exploitation.

By guiding stakeholders through a structured series of questions, the decision tree helps ensure that the decision to prioritise or defer mitigation efforts aligns with the organisation's overall goals and resource constraints. In goal-oriented approaches, this helps organisations avoid "one-size-fits-all" risk assessments and instead tailor their responses based on unique operational needs and priorities.

**2. Handling of Uncertain or Incomplete Data**

In IoT environments, there are often scenarios where precise quantitative data about vulnerabilities, likelihoods, or impacts are unavailable. SSVC's use of qualitative data offers a practical solution for addressing these uncertainties. By enabling decision-makers to categorise risks using qualitative descriptors, such as high, medium, or low impact,the SSVC framework facilitates risk prioritisation even when probabilistic data may be lacking or incomplete. This flexibility is particularly useful in dynamic IoT ecosystems, where new vulnerabilities may emerge faster than they can be quantified through traditional metrics.

For example, in a situation where a vulnerability is known to exist, but the exploitability is unclear due to a lack of historical data, SSVC allows stakeholders to make informed decisions based on qualitative assessments (e.g., whether the vulnerability is in a critical system or whether mitigations are already in place), rather than waiting for complete quantitative data.

**3. Enhanced Collaboration and Communication**

SSVC's decision tree structure simplifies the communication of risk decisions across multidisciplinary teams, including cybersecurity professionals, management, and other stakeholders. The step-by-step nature of the decision trees makes the reasoning behind prioritisation decisions more transparent and accessible, enabling better collaboration between technical and non-technical team members. In a goal-oriented approach, where aligning cybersecurity objectives with business and operational goals is crucial, the decision tree's clarity facilitates shared understanding and decision-making across different levels of the organisation.

By providing clear rationales for prioritising certain vulnerabilities over others, SSVC enhances the alignment between cybersecurity efforts and organisational goals, ensuring that resources are focused on the most critical vulnerabilities that pose the greatest threat to achieving those goals.

**4. Rapid and Adaptive Response to Emerging Threats**

Decision trees offer the advantage of enabling a more rapid and adaptive response to newly identified vulnerabilities. In fast-paced IoT environments, where new devices and technologies are frequently deployed, waiting for complete quantitative risk data may delay critical vulnerability mitigations. SSVC's decision trees provide an immediate framework for determining the severity of a vulnerability and the urgency of required action, allowing organisations to act quickly and adjust their strategies as new threats emerge.

For instance, if a new vulnerability in a widely used IoT device is discovered, the SSVC framework can quickly guide decision-makers through prioritisation steps, such as assessing whether the vulnerability affects critical operations or whether there are feasible mitigations in place. This agility is crucial in dynamic IoT environments, where the rapid identification and prioritisation of risks can prevent widespread system disruptions.

**5. Alignment with Existing Risk Management Standards**

SSVC's qualitative approach aligns well with existing cybersecurity standards, such as the **NIST Cybersecurity Framework** and **ISO/IEC 27001**, which also incorporate qualitative elements in their risk management processes. This makes SSVC compatible with widely used risk assessment methodologies, allowing organisations to integrate the SSVC decision tree approach into their broader cybersecurity management efforts seamlessly.

In a goal-oriented framework, this compatibility ensures that organisations can apply SSVC while still adhering to broader regulatory or compliance requirements, thus enhancing its practical application in both industry-standard and custom-tailored risk **management** strategies.

**SSVC's in a Goal-Oriented Approach**

SSVC's reliance on decision trees and qualitative data offers significant advantages for prioritising vulnerabilities in a goal-oriented approach. By allowing tailored decision-making, handling uncertainty, enhancing communication, and enabling rapid response, SSVC helps organisations align their cybersecurity efforts with operational goals more effectively. Its flexibility and adaptability make it a valuable tool for IoT environments where risks are constantly evolving, and quantitative data may not always be immediately available. The integration of SSVC into the proposed goal-oriented model strengthens the model's ability to assess and mitigate IoT cyber risks in a comprehensive and practical manner.

### *8.1 Simulation of the Goal-Oriented Approach*

Table 3 presents a critical simulation of the proposed goal-oriented approach to risk assessment, specifically focusing on the allocation of risk categories, principles, and their associated weightings. This table is an essential part of the research as it demonstrates how the proposed model can be applied in real-world settings for effective cyber risk assessment and mitigation in IoT environments. The following points elaborate on the significance and interpretation of Table 3:

**1. Categorisation of Risk Principles**

Table 3 is structured to categorise risks based on specific principles that reflect different dimensions of IoT security. Each risk category is allocated a unique code, which corresponds to a specific vulnerability or risk scenario identified in IoT systems. These principles encompass a wide range of security concerns, from technical vulnerabilities (e.g., device compromise) to broader strategic risks (e.g., reputational damage from data breaches).

The inclusion of these principles allows for a comprehensive assessment that goes beyond individual technical vulnerabilities, offering a more holistic view of the organisation's overall risk posture. This categorisation enables organisations to prioritise risks based on their relevance and severity in different IoT environments, such as smart cities, industrial IoT, or healthcare.

**2. Weighting System**

Each risk category is assigned a weight ranging from 0 to 3, depending on the likelihood and impact of the associated vulnerability or threat. The weighting is determined by evaluating the **exploitability** of the vulnerability and its potential to cause cascading failures across interconnected IoT devices and systems.

- **Weight 0** indicates minimal risk or low likelihood of exploitation.

- **Weight 1** indicates a moderate level of risk that requires monitoring but may not necessitate immediate intervention.
- **Weight 2** reflects a higher probability of exploitation with potentially significant consequences, warranting proactive risk mitigation.
- **Weight 3** indicates a critical risk that requires immediate action due to its potential to cause widespread disruptions or severe financial and operational damage.

The weighting system allows organisations to focus resources on the most pressing risks, enabling efficient allocation of security budgets and efforts to mitigate IoT-related cyber threats.

***Definition of Risk Exploitability and Weight Determination in Table 3***

**Risk exploitability** refers to the likelihood that a vulnerability or risk in an IoT system can be successfully exploited by a threat actor. In the context of IoT cybersecurity, exploitability is a crucial factor because not all identified vulnerabilities carry the same probability of being exploited. For instance, certain vulnerabilities may require advanced skills, specific conditions, or access to specific network segments to be exploited, while others can be easily exploited with widely available tools.

In this work, risk exploitability is determined based on several key factors:

1. **Access Complexity**: The ease or difficulty with which a threat actor can access the vulnerable component. This includes whether the vulnerability is exposed to the internet or resides behind secure layers like firewalls.
2. **Required Privileges**: The level of privileges or access control required to exploit the vulnerability. For example, a vulnerability that requires administrative privileges is typically harder to exploit than one that can be exploited by a standard user.
3. **Publicly Available Exploits**: Whether or not there are existing tools or scripts available to exploit the vulnerability. If an exploit is readily available and easy to use, the risk exploitability is higher.

4. **Attack Vector**: The means through which the attack is executed. For instance, vulnerabilities that can be exploited remotely over a network generally have higher exploitability than those requiring physical access to the device.
5. **Patch Availability and Mitigation**: Whether there are patches or mitigation strategies in place. A vulnerability with no available patch or limited mitigation options is more exploitable than one for which a patch exists and has been widely applied.

**Further Detail on the Weight Determination in Table 3**

The weights assigned to vulnerabilities in **Table 3** are based on an estimation of risk exploitability. Each risk category is evaluated using the factors mentioned above, and a numerical weight (ranging from 0 to 3) is assigned to represent the likelihood of exploitation. The weights correspond to the following levels of exploitability:

- **Weight 0 (Low Exploitability)**: This weight is assigned to vulnerabilities that have extremely low risk of being exploited. This may include vulnerabilities that require highly specialised skills, physical access to the device, or complex conditions that are unlikely to occur. For example, vulnerabilities that exist only in closed networks or require multiple layers of compromise to access would receive this weighting.
- **Weight 1 (Moderate Exploitability)**: Vulnerabilities with moderate risk of being exploited are assigned this weight. These might require some level of specialised knowledge or access but are feasible for an attacker to exploit under the right conditions. An example would be a vulnerability that requires privilege escalation within a network but does not have readily available public exploits.
- **Weight 2 (High Exploitability)**: This weight is assigned to vulnerabilities that are relatively easy to exploit and are likely to be targeted by attackers. These may involve publicly available exploits, easily accessible devices, or remote attack vectors. For instance, a vulnerability in an IoT device exposed to the internet without sufficient patching or protective measures would typically fall into this category.

- **Weight 3 (Critical Exploitability)**: Vulnerabilities that are extremely easy to exploit and carry severe consequences are assigned the highest weight. These include vulnerabilities for which widely used exploit kits are available, or where a remote attacker can easily gain control over a device or network segment. An example would be an unpatched zero-day vulnerability in an IoT system that is exposed to the public internet.

**Example of Weight Assignment in Table 3**

For example, in Table 3, if a particular IoT vulnerability exists in a publicly accessible smart device that requires minimal technical knowledge to exploit and has a widely available exploit tool, it would be assigned a weight of **3 (Critical Exploitability)**. In contrast, a vulnerability in a back-end server that requires significant expertise and internal network access might be assigned a weight of **1 (Moderate Exploitability)**.

The weighting is also influenced by the **dependency relationships** in the IoT system. If a vulnerability in one device can cause a cascading effect across multiple interconnected devices, its exploitability may be weighted higher due to the broader system-wide impact. Conversely, isolated vulnerabilities with limited impact are weighted lower.

**Integration with AI/ML Models**

The exploitability weights are incorporated into the proposed AI/ML-based risk assessment framework to dynamically adjust the risk scores of IoT components. The AI model processes these weights along with other input data (e.g., network traffic patterns, device telemetry) to produce real-time risk assessments. The use of exploitability weights ensures that the model prioritises the most severe and actionable risks, helping organisations focus their mitigation efforts on vulnerabilities that pose the highest threat.

By defining risk exploitability and assigning corresponding weights, this work introduces a structured and transparent method for prioritising vulnerabilities within IoT systems. This approach ensures that both easily exploitable and high-impact vulnerabilities receive the attention they warrant, while lower-risk issues are deprioritised. This allows for more efficient allocation of resources in mitigating IoT cyber risks, enhancing the overall security posture of IoT deployments.

**3. Vulnerability Exploitability and Decision Trees**

Table 3 links each risk category to the concept of **vulnerability exploitability**. By incorporating the **Stakeholder-Specific Vulnerability Categorisation (SSVC)** decision tree methodology, the table provides a dynamic assessment of risk scenarios. SSVC helps assess the decision points around patching vulnerabilities or applying other mitigation measures based on the risk's exploitability and the organisation's tolerance for risk.

This approach allows organisations to decide whether to accept, mitigate, or transfer a specific risk. For instance, for highly exploitable vulnerabilities that pose significant risk (assigned a weight of 3), the organisation might opt for immediate patching or enhanced monitoring. In contrast, less severe vulnerabilities with a lower weight might be mitigated over time or transferred via cyber insurance.

**4. Real-World Application of the Goal-Oriented Approach**

Table 3 demonstrates the practical applicability of the goal-oriented approach by simulating real-world scenarios where organisations need to assess and prioritise IoT-related risks. For example, in industrial IoT (IIoT) environments, the risk of a compromised sensor leading to downtime in a production line would be assigned a high weight due to the potential cascading impact on production processes. Similarly, in smart cities, the failure of IoT-connected traffic control systems could lead to significant public safety risks, requiring immediate mitigation strategies.

This simulation shows how the proposed framework can be applied across different domains and provides a clear roadmap for decision-makers to follow when assessing IoT risks. It enables them to make informed choices about where to allocate resources, how to prioritise risks, and which mitigation strategies (e.g., risk acceptance, transference, or mitigation) to adopt.

**5. Alignment with Global Standards**

The principles and weightings in Table 3 align with widely accepted cybersecurity frameworks, such as **ISO/IEC 27001**and **NIST Cybersecurity Framework**, making the table highly adaptable to different organisational contexts. The integration of global standards ensures that the approach is applicable not only to Cisco's operational environment but also to a wide range of industries and sectors, from healthcare to manufacturing and beyond.

**6. Future Refinements**

Table 3 provides a snapshot of the current state of the proposed goal-oriented approach, but it also points to potential areas for future refinement. For example, as new IoT vulnerabilities emerge or regulatory environments evolve, the weightings and principles in Table 3 can be updated to reflect the latest threat landscape. The flexibility of this framework allows it to remain relevant in the face of rapid technological change, ensuring that it can accommodate new developments in IoT security.

Table 3 illustrates the practical applicability and flexibility of the goal-oriented approach for IoT risk assessment. By categorising risks, applying weightings based on vulnerability exploitability, and integrating decision trees, this table offers a structured and actionable framework for organisations to assess and prioritise cyber risks in their IoT ecosystems. The alignment with global standards and the potential for future refinements ensure that the approach remains adaptable and generalisable to a wide range of operational contexts.

This case study brought forward the limitations of current exchange mechanisms on vulnerability data, with the main concern being around the fact that sharing exploitability data on vulnerabilities that have not been patched, exposes the risk of this data being intersected by hackers, enabling them to use exploits in real time before cybersecurity experts had sufficient time to patch the vulnerability. This is the main concern in terms of cyber risk, and this concern has been in circulation since 1990s. VEX is the latest attempt to resolve this long-term issue in cyber risk assessment of third-party risk.

### *8.2 CSAF/VEX and cyber insurance*

However, comparing our arguments of targeted data strategy for risk assessment, with the current model of cyber insurance works as a risk mitigation tool and covers the costs of losses caused by human malicious activity or natural disasters. In this context, many of the problems in the banking and financial sector and their failures of the past decade can be directly tied to model failure or overly optimistic judgments in the setting of assumptions or the parameterisation of a model. Now, a new public policy has emerged in which insurance companies act as clearing houses for information, integrate different security services and provide guidance on appropriate security investments to businesses seeking liability coverage (Allodi and Massacci 2017a). For example, new and traditional insurers can outsource important parts of the forensic investigation to different consultancies such as software or networking companies. However, recent research shows that this view of cyber insurance as a delegated policy tool has limitations in producing the anticipated coordination benefits and indeed may erode the aggregate level of security investment undertaken by targets in different insurance markets (Allodi and Massacci 2017b). These limitations are reflective of the previously discussed issue that insurance markets are lacking empirical data to construct actuarial tables. Thus, resulting with banks and insurers being unable to price IoT cyber risk with the same precision as in traditional insurance lines. While new and recent quantitative models partially address this issue, it may still be some time before these new approaches are widely adapted in the banking and cyber insurance sectors.

It should be recognised that IoT represents a huge opportunity for insurers to harness and understand cyber risks. IoT can thus represent a part of the solution to improved coverage and liability of non-tangible digital assets and to the dynamic nature of cyber-attacks. IoT can provide a part of the response to the general agreement that there is not enough data to understand the risks and reduce resource allocation problems arising from incomplete information regarding parties' actions (e.g. moral hazard) and characteristics (e.g. adverse selection). However, the analysis and correlation of large IoT data sources and new digital forensics and methods might sometimes be insufficient. For example, even though algorithms used to calculate cyber-risk metrics can analyse and correlate vast amounts of data, the methodologies that inform actuarial models may still struggle to make sense of and integrate the real-time information available from IoT devices.

Over-reliance on modelling in cyber insurance can also conceal difficult-to-detect processes, such as in the 'normalisation of deviance' case. The normalisation of deviance defines the processes that socially organise and systematically reproduce mistakes related to complex technological solutions. In this context, IoT can help by making use of data to increase transparency and predictability of such processes, understand the limitations of computational modelling and techniques and improve the assumptions that these models are based on. The constant inspection of granular IoT data and the possibility of sharing aggregates of IoT data and increasing transparency between parties can help insurers and re-insurers understand strict liability and its sharing across complex ecosystems. Parties can collaborate to prevent risks from cascading and to investigate possible "black swan" events (an event that is unprecedented and unpredicted) in relation to the use of digital devices, which are likely to increase in number, at least in the short term. Formal methods for trustworthiness assessment may then help inform insurance models for complex IoT ecosystems. Such developments would temper the proliferation of false beliefs due to over-reliance on the 'accuracy' of the outputs from computing models, which can lend an apparent objectivity to the results that can then justify inappropriate actions and policies.

From a risk management point of view, one important question is: what architectural improvements of a company's IT data system might increase resilience to cyber risk? We tend to think of cyber security as pertaining to IT companies, but digitalisation is currently extending well beyond the existing IT systems to manufacturing floors and other production activities that only a few people normally associate with IT data strategy. Thus, the boundaries between IT systems and operational activities, such as manufacturing, may not be obvious. Then, data processing is mobile and is constrained by the environment where the system operates. For example, when complex systems interact, it is very difficult to predict the system's behaviour and, in particular, the failure modes in operational conditions because of the emergent nature and the created feedback loops. This represents a particular challenge as, despite new investments in IoT and broad concerns with cyber risks, the manufacturing industry is still fragmented in its approach to managing cyber-related risks and having the organisational ownership to do so effectively (Buith 2016).

More general risks pertain to the vulnerability that IoT solutions currently have in relation to cyber-attacks and the capability of such solutions to establish and maintain different sorts of rights, such as the right to privacy. The relatively recent DDoS attacks that exploited simple, but poorly secured IoT end devices, such as baby monitors with immutable default passwords, show that: a). the model of low-cost, low-security IoT solutions is not sustainable, and that b). organisations and individuals need to protect themselves through collaborations, increased transparency, re-drawing of the current accountability and liability domains, and so forth. However, the mechanisms needed to implement these aspects need to work in a globalised context and across jurisdictions.

From a public policy point of view, insurers have become ‘de facto regulators’ by establishing a minimum-security level to gain cyber coverage. This argument emerges from research that links security controls and cyber insurance proposal forms. In this context, IoT can help shape public policies that are beneficial for the insurance sector and the society at large. For example, one important opportunity is represented by businesses using IoT to demonstrate their compliance with both national and international industry standards as well as internal policies. The challenges facing organisations in standards compliance for IoT systems are significant (Christensen et al. 2019).

However, insurers can design dynamic insurance policies that would not only reflect the changes in behaviour and characteristics of businesses and the contexts in which they operate but will also allow the creation of insured ecosystems where dynamic mechanisms such as double rewarding mechanisms and adaptive incentives can be operationalised. In such ecosystems, network-specific risks could be transferred between businesses and insurers (or re-insurers) in near real time, e.g. by using smart contracts. However, there are important limitations here as these systems operate in environments that are too complex to manage, cannot be rationally understood and pose specific moral questions such as those around privacy and data protection. The difficulty is that each of these aspects helps the system to evolve but it can also change the a priori allocation of risks. Nevertheless, dynamic risk assessment and dynamic insurance policies can improve some of the current challenges in cyber risk mentioned in this paper and can represent opportunities to improve the existing regulation, public policies and government interventions in the cyber insurance market.

Current developments in IoT ecosystems allow innovative ways to design cyber insurance services that would utilise IoT data to:

- Design of a tailored data strategy for IoT and cyber risk assessment;

- Mitigate risk management and facilitate new developments in this area, such as risk engineering;
- Increase transparency and predictability of the cyber insurance processes, including near real-time evidence-based explanations meant to increase trust and reduce risks;
- Increase the flexibility and adaptability of the current business environments, including the correlation of multi-model information such as risk, anomaly scores and liability;
- Enable co-evolution of systems, where learning and knowledge are distributed between the insurance company and the insured parties towards a still more efficient allocation of risk and responsibility, and
- Investigate the use of Smart Contracts to manage cyber risks within the insured environment.

The use of Smart Contracts raises one critical question on how empirical data can be collected and used with the dependency model to provide quantitative assessments. More comprehensive and systematic understanding of this question will arise when AI/ML technologies are migrated to the periphery of the internet and into local IoT networks. By integrating AI/ML in the dependency risk analytics, we can anticipate that real time intelligence data would enable dependency systems to recover and become more robust. AI/ML in the dependency risk analytics would also enable an understanding how and when compromises happen and enable systems to adapt and continue to operate safely and securely when they have been compromised.

### *8.3 Data Sources for Gathering Probabilistic Information in the Proposed Data Strategy*

Given the complexity and dynamic nature of IoT systems, the data sources included in the data strategy are drawn from diverse domains to ensure robust probabilistic risk assessments. The data strategy integrates real-time and historical data from multiple layers of IoT infrastructure, allowing for more accurate predictions and risk estimations. The key data sources are outlined below:

a. **Network Traffic Data**: One of the primary data sources is real-time network traffic data generated by IoT devices and networks. This includes data on the types, frequency, and volume of communications between IoT devices, servers, and external systems. Analysing network traffic enables the identification of anomalies, such as unusual patterns of data transmission, which could indicate cyber threats like botnet activity, distributed denial-of-service (DDoS) attacks, or data exfiltration. Tools such as **intrusion detection systems (IDS)** and **network monitoring platforms** provide raw traffic data, which is processed to extract probabilistic insights on attack likelihood and impact.

b. **Device Telemetry Data**: Telemetry data from IoT devices includes metrics related to device health, performance, and operational status. This data is crucial for understanding the normal operational baseline of IoT devices and detecting deviations that could signal a cyber-attack or device malfunction. For example, abnormal energy consumption or processing delays could indicate that a device is compromised. Telemetry data is gathered through **device management platforms** and **cloud-based IoT hubs** that aggregate information from multiple devices in real-time.

c. **Incident Logs and Historical Attack Data**: Historical incident logs from security breaches, cyber-attacks, and device failures serve as a valuable source of probabilistic information. These logs provide insights into attack vectors, timelines, and vulnerabilities exploited in past incidents. By examining patterns in historical data, the proposed model can estimate the likelihood of future attacks. Data sources such as **Security Information and Event Management (SIEM) systems**, **firewall logs**, and **threat intelligence platforms** are critical for gathering and analysing this historical information.

d. **Vulnerability Databases**: Publicly available vulnerability databases, such as the **National Vulnerability Database (NVD)** and the **Common Vulnerability Scoring System (CVSS)**, provide critical data on known vulnerabilities in IoT devices, software, and protocols. These databases are constantly updated with information on newly discovered vulnerabilities, enabling organisations to assess the likelihood of exploitation based on the severity and type of vulnerability. These data sources are used to quantify the risk of unpatched vulnerabilities being exploited in a probabilistic framework.

e. **Threat Intelligence Feeds**: External threat intelligence feeds, such as those provided by **commercial security vendors** or open-source platforms, offer real-time information on emerging cyber threats, attack techniques, and indicators of compromise (IoCs). These feeds are crucial for staying updated on new attack patterns targeting IoT ecosystems. Integrating threat intelligence feeds allows the model to dynamically adjust its risk estimates based on real-time threat levels. Sources include **MITRE ATT&CK**, **FireEye**, **Palo Alto Networks**, and **IBM X-Force**.

f. **IoT-Specific Sensor Data**: In environments where IoT devices are deeply integrated with physical systems, such as smart cities, industrial automation, and healthcare, sensor data plays a key role in identifying risk scenarios. For example, sensor data from smart meters, connected vehicles, or industrial equipment can indicate when devices are behaving abnormally, allowing for early detection of potential threats. **SCADA systems** and **IoT platforms** typically aggregate this data, which can then be used to estimate the likelihood of equipment failure or cyber-physical attacks.

g. **Third-Party Security Audits and Compliance Reports**: Organisations often conduct security audits and assessments of their IoT infrastructure to ensure compliance with industry standards such as **ISO/IEC 27001** or **NIST Cybersecurity Framework**. These audits provide valuable data on system weaknesses, compliance gaps, and potential threats. The results of these audits are included in the probabilistic model to help determine the organisation's overall cyber risk posture and identify areas where additional mitigation measures are necessary.

h. **Cyber Insurance Claims Data**: Data from cyber insurance claims offers a unique perspective on the financial impact and frequency of cyber incidents. Claims data can provide insights into the types of attacks that lead to significant financial losses and the effectiveness of risk transfer mechanisms, such as insurance. This data can be aggregated from **insurance companies**, **brokers**, and **industry reports**, and can help calibrate the financial risk models in the proposed strategy.

**Combining Data Sources for Enhanced Probabilistic Assessment**

The proposed data strategy leverages these multiple data sources to create a comprehensive dataset for probabilistic analysis. The data is processed using **machine learning algorithms** and **statistical models** to estimate the likelihood of various cyber threats and their potential impact. By integrating real-time data with historical trends and external threat intelligence, the model can dynamically adjust risk estimates and provide a more accurate and proactive assessment of IoT cyber risks.

This approach ensures that organisations can move beyond static, qualitative assessments and rely on data-driven, probabilistic insights to inform their IoT security strategies. The inclusion of diverse data sources also makes the model adaptable to different IoT environments and threat landscapes.

### *8.4 Adoption and Impact of VEX on Third-Party Cyber Risk Assessment*

**Current Level of VEX Adoption in the Cybersecurity Community**

The **Vulnerability Exploitability eXchange (VEX)** has emerged as a relatively new but increasingly important tool in the cybersecurity community for improving the precision of vulnerability management and third-party risk assessments. Developed in response to the long-standing challenge of assessing the exploitability of known vulnerabilities in real-time, VEX is being gradually adopted, especially in industries where supply chain security and third-party risk are critical.

As of the time of writing, VEX adoption is still in its early stages but gaining traction, particularly in the following areas:

1. **Adoption in Software Supply Chain Security**: With growing concerns over supply chain vulnerabilities, VEX is being increasingly integrated into **Software Bill of Materials (SBOM)** frameworks to provide more granular and timely information about which vulnerabilities in a software component are exploitable. The U.S. **National Telecommunications and Information Administration (NTIA)** and **National Institute of**

**Standards and Technology (NIST)** have both recognised the importance of VEX in their cybersecurity guidance, and its use is being advocated in sectors such as healthcare, critical infrastructure, and defence, where software supply chain risks are particularly high.

2. **Industry Adoption**: Several major vendors and cybersecurity providers, including those in cloud services and IT management, are beginning to incorporate VEX profiles into their vulnerability management tools. For example, leading providers of vulnerability assessment platforms and risk management solutions are adding support for VEX to improve the precision of vulnerability prioritisation, particularly when assessing the security of third-party software and services.
3. **Regulatory Push for Adoption**: The inclusion of VEX in key U.S. government initiatives, such as **Executive Order 14028** on improving national cybersecurity, is driving broader adoption across critical industries. The Executive Order calls for improved transparency in software components through SBOMs, with VEX providing essential details on the real-world exploitability of vulnerabilities. This regulatory push is influencing sectors such as energy, finance, and telecommunications to adopt VEX as part of their vulnerability management and compliance efforts.

**Assessment of VEX's Impact on Improving Third-Party Cyber Risk Assessment**

Although VEX is relatively new, initial assessments of its impact suggest that it has the potential to significantly improve the way third-party cyber risks are assessed and managed. Some of the key benefits and emerging impacts of VEX on third-party risk assessments are as follows:

1. **Precision in Vulnerability Prioritisation**: One of the primary advantages of VEX is that it allows organisations to focus on vulnerabilities that are truly exploitable, rather than wasting resources on vulnerabilities that may not pose a real threat. In third-party risk assessments, this added precision helps organisations more effectively evaluate the security posture of their vendors and partners by identifying which vulnerabilities in third-party software are exploitable within their operational environment. This shift reduces false positives and minimises the burden of patching non-critical vulnerabilities.

2. **Reduction of Patch Fatigue**: Third-party vendors often release patches for vulnerabilities that may not be exploitable in all environments. With VEX, organisations can more effectively prioritise which patches to apply based on actual exploitability data, reducing "patch fatigue" among IT and security teams. This has been especially impactful in environments with extensive vendor relationships and dependencies, such as **cloud computing** and **SaaS providers**, where constant updates and patches can be overwhelming.
3. **Improved Supply Chain Risk Management**: VEX improves transparency across the software supply chain by providing explicit, machine-readable information about whether a vulnerability in a software component is exploitable. This enhanced visibility allows organisations to better manage risks across their third-party ecosystem, which is crucial for mitigating supply chain attacks such as those seen in incidents like **SolarWinds** or **Log4j**. Initial industry feedback indicates that organisations using VEX-enabled SBOMs can more effectively respond to vulnerability disclosures and reduce their exposure to third-party risks.
4. **Integration with Cyber Insurance Models**: Another emerging impact of VEX is its potential role in cyber insurance underwriting. By providing more precise data on the exploitability of vulnerabilities in third-party software, VEX enables insurance providers to better assess the cyber risk posture of insured parties. This could lead to more accurate pricing of cyber insurance policies and incentivise better vulnerability management practices across the supply chain.
5. **Enhanced Regulatory Compliance**: VEX's machine-readable format aligns well with the increasing demands for transparency and accountability in cybersecurity regulations. In industries where compliance with security standards is essential (e.g., healthcare, financial services), VEX can help organisations demonstrate that they are addressing truly exploitable vulnerabilities, thereby enhancing compliance with frameworks like **NIST 800-53** or **ISO/IEC 27001**. This impact is particularly important in the context of third-party risk management, where regulatory bodies are increasingly requiring organisations to take greater responsibility for the security of their entire supply chain.

**Challenges to VEX Adoption and Its Future Prospects**

While VEX shows great promise, its adoption is still facing several challenges:

- **Standardisation and Interoperability**: Although VEX is being promoted as a standard, different vendors may interpret or implement it differently, leading to issues with interoperability between tools and platforms. Efforts are underway to establish more unified standards and guidelines to streamline VEX's use across the industry.
- **Education and Awareness**: Many organisations are still unfamiliar with VEX and its benefits. As with any new standard, significant efforts are required to educate both vendors and users on how to implement and leverage VEX for more effective vulnerability management.

Nevertheless, with the continued regulatory push for software transparency, the increasing complexity of supply chain attacks, and the growing emphasis on third-party risk management, VEX is likely to see broader adoption in the coming years. As more organisations incorporate VEX into their SBOMs and vulnerability management processes, its impact on improving third-party cyber risk assessments will become more evident, contributing to a more resilient cybersecurity ecosystem.

# 9 Discussion

## *9.1 Generalisation of the Proposed Model to Real-World Scenarios*

While the BoT-IoT dataset provides a valuable basis for developing and testing the proposed model for IoT cyber risk assessment, it is essential to acknowledge that the model is designed to be generalisable to a wide range of real-world IoT environments. The challenges posed by the BoT-IoT dataset, such as botnet attacks, DDoS scenarios, and other cyber threats, reflect a subset of the broader set of cyber risks faced by IoT systems. However, the proposed model is not limited to the specific characteristics of this dataset and can be applied to other real-world scenarios in various IoT ecosystems.

The key features of the model that enable its generalisation include:

a. **Dependency Modelling**: The use of dependency modelling in the proposed approach is highly flexible and can accommodate different types of IoT systems, from smart homes to industrial IoT (IIoT) environments. By focusing on the interactions and interdependencies between IoT components, such as devices, networks, and data flows,the model can be adapted to capture cyber risks in complex, real-world systems where threats arise from diverse sources. This is particularly important in real-world deployments where different vendors, protocols, and device architectures coexist, creating unique vulnerabilities.

b. **Scalability to Heterogeneous IoT Environments**: IoT systems in real-world scenarios often involve heterogeneous devices with varying levels of security and functionality. The proposed model, by abstracting key risk factors such as network topology, communication protocols, and device types, is well-suited for application in environments where these elements differ significantly. For example, in smart city infrastructure, the same dependency-based risk assessment methodology can be used to assess risks in traffic management systems, connected energy grids, or public safety networks, despite the differences in the nature of devices and data flows involved.

c. **Inclusion of Multiple Attack Vectors**: The model is adaptable to multiple attack vectors, beyond the botnet attacks simulated in the BoT-IoT dataset. In real-world applications, IoT systems are susceptible to a wide range of attacks, such as malware infections, ransomware, zero-day vulnerabilities, and data breaches. The proposed model's flexible risk estimation framework can be extended to incorporate new attack vectors as they emerge, ensuring that it remains relevant in the ever-evolving threat landscape.

d. **Applicability to Different IoT Domains**: Although this study focused on a dataset tailored to a specific subset of IoT risks, the model can be applied to other critical domains such as healthcare, industrial automation, and connected transportation. For instance, in healthcare IoT, where safety and data integrity are paramount, the same risk estimation principles can be applied to assess the cyber risks posed by compromised medical devices or the failure of patient-monitoring systems. Similarly, in IIoT environments, the model can be adapted to assess risks in the context of supply chain disruptions or physical damage caused by cyber-physical system failures.

e. **Adaptability to Emerging IoT Cybersecurity Standards**: The model's framework can also be integrated with evolving IoT cybersecurity standards and regulatory frameworks. For example, the NIST Cybersecurity Framework for IoT and ISO/IEC 27001 standards provide guidelines that can be mapped onto the proposed model's structure, ensuring that it remains aligned with industry best practices and can be easily adopted by organisations seeking compliance with these standards.

**Real-World Application Examples**

To illustrate the potential for generalisation, consider the following real-world examples where the model could be applied:

- **Smart Cities**: The proposed model could assess the risks in smart traffic systems, where compromised IoT devices like traffic lights or surveillance cameras could lead to large-scale disruptions.
- **Healthcare IoT**: In a hospital setting, the model could help assess the risk of data breaches in IoT-connected medical devices, such as insulin pumps or heart monitors, which could have serious implications for patient safety.
- **Industrial IoT**: The model could be used in a factory setting to assess risks to connected machinery, where a failure in one system could cascade through others, disrupting the entire production line.

In these cases, the dependency-based risk assessment and mitigation strategies proposed in this model would be applicable even when faced with varying device types, communication protocols, and risk profiles.

While the BoT-IoT dataset served as a valuable starting point for evaluating the model's performance, the model is inherently designed to be adaptable to real-world IoT environments that extend beyond this dataset. Its flexibility, scalability, and ability to accommodate new attack vectors make it highly generalisable across a variety of sectors and use cases. Future work will focus on further testing and refinement of the model in live IoT environments to validate its effectiveness in mitigating cyber risks across different domains.

### *9.2 Generalisability of Findings on IoT and Risk Transference*

While this research benefited from substantial input and access to Cisco's cyber risk management environment, the findings on IoT cyber risks and risk transference are designed to be generalisable to a wide range of organisations beyond Cisco. Several factors support this generalisability:

a. **Common IoT Risk Factors**: The IoT-specific risks identified in this research, such as interoperability challenges, cascading failures, and vulnerabilities in connected devices,are common across many industries and sectors. These risks are not unique to Cisco's operational environment but reflect broader trends observed in IoT ecosystems worldwide, such as in healthcare, industrial automation, and smart cities. Therefore, the risk transference strategies proposed in this research can be applied to any organisation facing similar challenges in managing interconnected IoT devices.

b. **Industry-Agnostic Risk Transference Strategies**: The concept of risk transference, particularly through mechanisms like **cyber insurance**, is not exclusive to Cisco's environment. Risk transference frameworks, such as the ones discussed in this study, apply universally to organisations that seek to mitigate the financial and operational risks posed by IoT-related cyber threats. For instance, cyber insurance policies, third-party liability agreements, and outsourcing of security functions are strategies used across multiple industries to shift risk exposure. As such, the recommendations made in this research can be adopted by a variety of organisations seeking to develop robust IoT risk management strategies.

c. **Framework Applicability Across Diverse IoT Environments**: The dependency modelling approach used in this research is flexible and adaptable, making it applicable to different types of IoT deployments and architectures beyond Cisco. By focusing on interdependencies between IoT devices, data flows, and cyber-physical systems, this model can be tailored to various operational environments, such as connected manufacturing lines, smart healthcare systems, and autonomous vehicle networks. The IoT risks and mitigation strategies discussed in this research therefore extend well beyond Cisco's specific use case and are relevant for organisations with similar IoT-driven infrastructures.

d. **Global Cybersecurity Standards and Practices**: The findings of this research are grounded in widely accepted cybersecurity standards, such as the **NIST Cybersecurity Framework** and **ISO/IEC 27001**, which are globally applicable and not specific to Cisco's internal practices. These standards promote best practices in cyber risk management and can be adapted by organisations of all sizes and sectors. The alignment of this study's findings with these international frameworks further reinforces the generalisability of the results across different organisational contexts.

e. **Broader Input from Multiple Experts**: Although Cisco provided valuable insights, the research also incorporated feedback from a variety of cybersecurity experts, representing different specialisations beyond Cisco's operational environment. This helped ensure that the findings, particularly those related to risk transference and IoT cyber risks, were not limited by the perspective of a single organisation. The collaboration with experts in IoT security, network vulnerabilities, and cyber risk frameworks has made the findings more applicable to a broader range of organisations.

The findings of this research are designed to be applicable to organisations across various industries and sectors. The identified IoT risks, dependency modelling, and risk transference strategies reflect global trends and are supported by widely accepted cybersecurity standards, making them highly relevant to organisations seeking to mitigate IoT risks in diverse operational environments. Future research could explore additional case studies in different sectors to further validate the generalisability of these findings.

### *9.3 Ensuring Explainability and Transparency in AI/ML for Dependency Risk Analysis*

The integration of AI/ML in dependency risk analysis offers significant advantages, particularly in identifying complex relationships and patterns that may not be apparent through traditional risk assessment methods. However, one of the key challenges associated with AI/ML in cybersecurity, and specifically in IoT dependency risk analysis, is ensuring that the decision-making process remains explainable and transparent. Stakeholders, including cybersecurity professionals and decision-makers, must be able to understand how AI/ML systems arrive at their conclusions, especially in high-stakes environments like IoT, where decisions may affect safety and critical operations.

To address these challenges, several strategies and best practices can be implemented to improve explainability and transparency:

**1. Use of Explainable AI (XAI) Techniques**

Explainable AI (XAI) is an emerging field that focuses on making AI and ML models more interpretable without sacrificing performance. XAI techniques ensure that decisions made by AI models can be traced back to understandable factors. When applying XAI to dependency risk analysis in IoT systems, the following methods can be utilised:

- **Feature Importance**: In dependency risk analysis, ML models typically analyse multiple features (e.g., network traffic, device telemetry, historical attack data). Feature importance techniques, such as **SHAP (SHapley Additive exPlanations)** or **LIME (Local Interpretable Model-agnostic Explanations)**, can help explain which features had the most significant impact on the model's decision. For example, if the model flags a specific IoT device as a high-risk point in the network, the feature importance analysis can show whether this is due to abnormal data traffic, historical vulnerabilities, or dependency with critical systems.

- **Model-Agnostic Approaches**: These approaches enable the analysis of complex models (e.g., deep learning or ensemble methods) by generating interpretable approximations. LIME, for instance, creates a locally interpretable linear model around the prediction, helping to clarify how a black-box model arrived at a specific conclusion regarding risk dependencies in IoT systems.

**2. Interpretable ML Models**

In some cases, using inherently interpretable models can be an effective way to ensure transparency. While deep learning models or complex neural networks may offer high accuracy, they can be difficult to explain. Instead, opting for more interpretable models, such as **decision trees, random forests, or logistic regression**, can provide a clearer path from input data to decision output. These models, though potentially less complex, offer higher explainability in decision-making processes for dependency analysis.

For example, decision trees, which mimic human decision-making logic, can be used to illustrate how specific vulnerabilities or IoT device dependencies lead to a higher overall risk score. Each branching point in the tree reflects a critical decision, making the process transparent and easy to follow.

**3. Traceability and Auditability**

For any AI/ML-based risk assessment model, it is crucial to ensure **traceability** and **auditability**. This involves maintaining logs and records that track every decision made by the AI/ML system. These records allow cybersecurity analysts to trace the steps leading to a specific risk prediction, ensuring that every decision can be reviewed and validated post-decision.

- **Traceable Workflows**: Implementing a workflow that tracks every action, from data ingestion to model training and prediction, ensures that the decision-making process remains transparent. These workflows can include detailed documentation of which models were used, the data they were trained on, and how predictions evolved over time.

- **Model Audits**: Regular audits of the AI/ML models used for dependency risk analysis should be conducted to ensure their outputs remain aligned with real-world data and organisational goals. This process can involve reviewing how new data affects predictions and making adjustments to the models as necessary to maintain accuracy and transparency.

**4. Human-in-the-Loop Systems**

To maintain a high level of trust and transparency, many AI/ML systems in cybersecurity integrate a **human-in-the-loop** approach. This method involves human analysts in critical decision points, allowing them to validate, refine, or override AI/ML-generated risk assessments. This hybrid approach combines the efficiency of automated analysis with the intuition and domain expertise of human cybersecurity professionals.

In dependency risk analysis, human experts can review key AI-driven decisions, particularly in cases where the model's output is uncertain or where the risks involve critical infrastructure. By keeping humans engaged in the decision-making process, organisations can ensure that all decisions are explainable and supported by both machine intelligence and human judgment.

**5. Transparency in Data Sources and Model Inputs**

Ensuring transparency begins with the **data inputs** used to train and operate AI/ML models. The types of data used for dependency risk analysis, such as network traffic logs, vulnerability databases, and IoT telemetry data,should be well-documented and made available for inspection. This transparency ensures that stakeholders understand the source of the model's knowledge and can assess whether the data used is relevant, up-to-date, and of sufficient quality.

- **Data Provenance**: Documenting the origin of data, such as which vulnerability feeds or IoT device logs were used, ensures that all stakeholders can understand the foundation of the AI/ML model's decisions. This is particularly important when models incorporate third-party data, as the reliability of this data directly affects the model's output.
- **Open Datasets**: Wherever possible, using open datasets or sharing anonymised data sources improves transparency and allows third parties to verify the models. For instance, the use of datasets like **BoT-IoT** or open vulnerability databases ensures that the data sources can be scrutinised and understood by a wider audience.

**6. Clear Risk Reporting and Visualisation**

One of the key ways to ensure transparency in AI/ML models is to provide clear and understandable visualisations of the model's risk assessments. **Dashboards** that visualise key risk metrics, such as the likelihood of exploitability, device dependencies, and impact of failures,make AI/ML-driven decisions more interpretable for non-experts. These dashboards should offer:

- Visual representations of how different IoT devices are interconnected,
- Risk scores for individual devices or systems,
- Explanations of how changes in dependencies influence overall system risk.

By providing clear and detailed risk visualisations, organisations can help all stakeholders, from technical staff to decision-makers, understand and trust the AI/ML outputs.

**7. Model Testing and Validation**

To ensure that AI/ML models used in dependency risk analysis are both accurate and explainable, it is essential to rigorously test and validate the models against real-world data. This process involves:

- **Benchmarking against known outcomes**: Testing the model against historical data to ensure it makes accurate predictions based on past incidents.
- **Cross-validation**: Ensuring that the model's predictions generalise across different datasets and scenarios, helping to confirm that it is robust and transparent in different IoT environments.
- **Regular updates**: Continuously updating the models with new data and testing their performance ensures that the AI/ML models stay relevant and interpretable as new IoT vulnerabilities emerge.

Ensuring explainability and transparency in AI/ML models for dependency risk analysis is crucial for maintaining trust in the decision-making process. By leveraging XAI techniques, using interpretable models, implementing human-in-the-loop systems, ensuring transparency in data sources, and providing clear risk visualisations, organisations can ensure that the AI/ML-driven decisions are understandable and actionable. These steps allow stakeholders to gain insight into how AI/ML models arrive at their conclusions, building confidence in the overall risk assessment process.

## 10 Conclusion

The findings of this research emphasise the critical need for a comprehensive risk assessment framework tailored to the unique challenges of IoT environments. Through the development of a dependency-based cyber risk model, this study highlights the significance of interdependencies among IoT components in understanding and mitigating cyber risks. The integration of AI/ML techniques enhances the model's adaptability, offering dynamic risk assessments based on real-time data, while ensuring transparency through explainable AI (XAI) methodologies. Furthermore, the exploration of risk transference strategies such as cyber insurance demonstrates practical approaches for mitigating financial and operational impacts. By empirically validating the model using the BoT-IoT dataset, the research provides a robust tool that can be generalised across diverse IoT domains, contributing to the development of a more secure and resilient IoT ecosystem. These contributions lay the groundwork for future advancements in IoT cybersecurity, particularly in refining AI-driven solutions and addressing the evolving landscape of IoT threats.

This article reviews existing literature on emerging trends in IoT risk assessment, including the emergence of the Software Bill of Materials (SBOM), the Vulnerability Exploitability eXchange (VEX) and the Common Security Advisory Framework (CSAF). Although there is a wealth of research on the values of a Bill of Materials in cyber risk assessment, there is very little work on the software components used in low-cost IoT devices. The Software Bill of Materials (SBOM) was developed to address this issue, but analysing vulnerabilities from SBOMs usually results in a serious workload for cybersecurity professionals. Much of this process can be automated, and in this article, we reviewed some potential solutions for such automation. The article proposes a dependency model based on the goal-oriented approach, designed to be compliant and supportive of the new Stakeholder-Specific Vulnerability Categorization (SSVC) based on decision trees.

Through reviewing existing risk methods, in this paper, we determined that the existing models, individually, do not provide solutions for impact estimation of IoT cyber risk in autonomous systems. This research builds upon integrating the existing models and presents a unifying model incorporating IoT cyber risks in the impact estimation. The challenge in testing and verifying this new 'combined/unified model' and ensuring that the new model addresses the IoT context is resolved with dependency modelling. To test and verify the new model, we designed dependency relationships. In **FIGURE 2** we describe what the connections (arrows) mean, and how dependencies are expressed, and we give a description of dependency presented in the paper. The proposed cyber risk assessment with a unified model and dependency modelling is designed to estimate IoT risks, the impacts caused by failures that cascade and aggravate the impacts from one affected system or component to another. Since IoT risks are decentralised through networked objects, such risk is often invisible in the risk assessments with methodologies designed for general cyber risk assessment. Our approach is designed to advance the IoT risk assessment discipline. It considers the dependencies in 'no-win' scenarios, where each scenario leads to a risk we cannot protect from. The dependency states we considered are risk acceptance and risk transference.

This paper provides an overview of the current IoT cyber risk assessment research, with specific new models on this topic, such as dependency modelling and cyber risk mitigation and transference strategies (e.g., cyber risk insurance). The paper refers to several models and risk assessment articles and technical publications that have emerged recently in the research literature. This research is important because it covers the lack of specific standards to govern the assessment of IoT cyber risk. The paper contributes to the current efforts to advance the understanding of risk in IoT systems and to produce a standardised design and a holistic approach for IoT risk assessment. Although our unified approach though dependency modelling does not resolve all the issues we identified in this article, this work represents an important step forward for the discipline.

In this article, what we argue is that what is really needed to improve cyber risk assessment, are rigorous mathematical and verifiable experimental results. Hence, we have conducted this work in collaboration with the FAIR Institute (Factor Analysis of Information Risk), the North Carolina Chapter of FAIR (FAIR 2023), and we applied the FAIR by design principles (Wilkinson et al. 2016). We are one of the leading protagonists in using quantitative methods, instead of the currently used qualitative and hybrid cyber risk assessment methods. That, however, needs confidence intervals, time bound ranges, frequency, distribution, and many other data inputs that we currently do not have. We argue that when AI/ML can be shifted to the IoT devices operating at the edge of the network, this data could be possible to collect autonomously, and that would enable moving on from qualitative and hybrid assessment, into a qualitative cyber risk assessment that uses rigorous mathematical reasoning to deliver verifiable experimental results.

### *10.1 Limitations of this study and opportunities for further research:*

Using the new design of a unified and holistic model for IoT risk assessment and risk management without the required probabilistic risk data remains a challenge. To test and verify the new design, this study applied the case study research method, conducted individual interviews, and conducted workshops with Cisco experts in cybersecurity. To prove the new design further, we also conducted 6-month long action research with Cisco and recorded the performance of the design, then made iterative improvements to ensure functionality in different real-world environments. The solution presented in this paper is the final version of the new design; multiple versions were tested in the process. However, most failed in the application stage, usually because they have proven challenging to implement or even to understand by experts who didn't build the method. The selection criteria were based on the experts' ability to understand and use the new process. The rationale behind this was that if a cybersecurity expert cannot use the system, it would be almost impossible to train a non-expert to use the system, and occasionally, we require different expertise in the risk assessment process.

Prior to attempting to use the new unified/holistic model, appropriate data strategies should be developed that would enable the collection of probabilistic data. Given the lack of standards and regulations on developing the required data strategies (for IoT cyber risk), it seems that private sector is leading these efforts rather than national statistical offices. However, without standards, regulations, and policies in place, it is hard to see how individual data strategies of private companies could be synchronised to enable sufficient probabilistic data for a comprehensive understanding of IoT cyber risks. To promote advancements in collection of probabilistic data through appropriate data strategies, further research should focus on the combination of regulations, standards, and policies on data collection of IoT risk, artificial intelligence for data collection from IoT sensor networks, IoT data safety, IoT cyber security and data collection from IoT equipment, along with ethics of machine learning in IoT cyber risk data collection. Interdisciplinary research such as this would benefit the process of identify a dynamic and self-adapting system supported with AI/ML and real-time intelligence for predictive cyber risk analytics for edge computing. The current state of our knowledge on this topic is that 'overcoming the alleged limitation of model-centric AI may well require paying extra attention to the alternative data-centric approach' (Hamid 2022). In other words, the current position in existing literature is that to resolve the problem with absence of probabilistic data, we need to look at how we structure our data strategies, and then consider the algorithms we use, in combination with the data states and properties. We must note that applying the proposed holistic model for IoT risk assessment and risk management is a challenge in the absence of relevant probabilistic data. This in turn requires developing appropriate data strategies to enable the collection and processing of required probabilistic data. This links to the currently increasing demands on developing data-centric approaches in the development of AI technologies which, with machine learning (ML) techniques, forward to the development of the IoT. This would enhance our capacity for a comprehensive understanding of the opportunities and threats that arise when edge computing nodes are deployed, and when AI/ML technologies are migrated to the periphery of the internet and into local IoT networks.